\documentclass[titlepage,12pt]{utarticle}
\pdfoutput=1
\usepackage{amssymb, amsmath, amsopn, amsthm,amscd,amsfonts}
\usepackage{color}
\usepackage{latexsym}
\usepackage{ifthen}
\usepackage{epsfig}
\usepackage{graphicx}
\usepackage{graphics}
\usepackage{mathtools}
\usepackage{bm,longtable,enumerate}
\numberwithin{equation}{section}
\newcommand{\tr}{{\rm tr}}

\newcommand{\wt}{\widetilde}
\newcommand{\wh}{\widehat}

\newcommand{\cA}{{\cal A}}

\newcommand{\eqn}{\begin{eqnarray}}
\newcommand{\eqnx}{\end{eqnarray}}

\def\beq{\begin{equation}}
\def\eeq{\end{equation}}
\def\beqa{\begin{eqnarray}}
\def\eeqa{\end{eqnarray}}
\def\scst {\scriptscriptstyle}
\def\tr{\mathop{\rm tr}\nolimits}

\def\matt[#1,#2,#3,#4]{\left(%
\begin{array}{cc} #1 & #2 \\ #3 & #4 \end{array} \right)}

\def\v2#1{\vv2[#1]}
\def\vv2[#1,#2]{\left(%
{#1 \atop #2}\right)}

\def\beq{\begin{equation}}
\def\eeq{\end{equation}}
\def\beqa{\begin{eqnarray}}
\def\eeqa{\end{eqnarray}}
\def\scst{\scriptscriptstyle}

\begin{document}
\preprint{DCPT-12/37 \\
DIAS-STP-12-09 \\
}

\title{{\bf Production of negative parity baryons \vskip 0.3cm in the holographic Sakai-Sugimoto model}}

\author{{\sc Alfonso Ballon-Bayona}
 \address{
  	Department of Mathematical Sciences, \\
         Durham University, \\
         South Road, \\
         Durham DH1 3LE, \\
         United Kingdom.\\[0.1cm]
         {\tt c.a.m.ballonbayona@durham.ac.uk}
         }
\\ {\sc Henrique Boschi-Filho $^{\rm b}$, Nelson R.~F.~Braga}
 \address{
 Instituto de F{\'i}sica, \\
 Universidade Federal do Rio de Janeiro,\\
         Caixa Postal 68528,\\
         21941-972 Rio de Janeiro, RJ, Brasil.\\[0.1cm]
         {\tt boschi@if.ufrj.br},\\
         {\tt braga@if.ufrj.br},\\
         {\tt mtorres@if.ufrj.br}
         	        } ,
{\sc Matthias Ihl}
 \address{
        School of Theoretical Physics, \\
        Dublin Institute for Advanced Studies, \\
        10 Burlington Rd, \\
        Dublin 4, Ireland. \\[0.1cm]
        {\tt msihl@stp.dias.ie}
  }
\\ {\sc Marcus A.~C.~Torres} $^{\rm b}$
}

\Abstract{We extend our investigation of resonance production in the Sakai-Sugimoto model to the case of negative
 parity baryon resonances. Using holographic techniques we extract the generalized Dirac and Pauli baryon form factors as
well as the helicity amplitudes for these baryonic states. Identifying the first negative parity resonance with the experimentally observed $S_{11}(1535)$,
we find reasonable agreement with experimental data from the JLab-CLAS collaboration. We also estimate the contribution of negative parity baryons to the proton structure functions.}

\maketitle
\tableofcontents
\newpage

\section{Introduction}

In the past decade, gauge/gravity dualities inspired by the original Maldacena conjecture \cite{Maldacena:1997re} have been successfully applied to a wide range of
problems in Quantum Chromodynamics (QCD) as well as condensed matter theory. For QCD, two classes of models are of particular interest: The phenomenological bottom-up approach based on
five-dimensional effective actions, and the more stringent top-down models based on compactifications of ten-dimensional string theories\footnote{Recommended reviews are  \cite{hep-th/0610135,arXiv:0908.0312,arXiv:1001.1978,arXiv:1006.5461} for the
bottom-up approach and \cite{Peeters:2007ab,Erdmenger:2007cm,Gubser:2009md,arXiv:1101.0618} for the top-down approach.}. A prominent and widely studied model of the latter class clearly is the Sakai-Sugimoto model \cite{Sakai:2004cn, Sakai:2005yt}. Its popularity can be
attributed to its close resemblance to large-$N_c$ QCD, its computational simplicity and the fact that it provides a geometric model that allows
to study both the confinement/deconfinement transition (at finite temperature) and chiral symmetry breaking in the same unified framework.

\medskip

Holographic techniques have also emerged as a very fruitful complementary tool for studying hadronic scattering in QCD
where non-perturbative effects become important, namely in the regime of low momentum
transfer ($\sqrt{q^2}$ lower than a few GeVs).
In this paper we investigate the production of negative parity baryon resonances in
proton electromagnetic scattering within the framework of the Sakai-Sugimoto model.
This is a natural continuation of our previous work on positive parity
baryonic resonances \cite{Bayona:2011xj}.
Production of baryonic resonances is a very important and timely problem in hadronic physics for the following reasons:
i) Many baryonic resonances are excited nucleon states ($N^\ast$) and their structure is relevant to
understand the physics of quark confinement, ii) there is a huge
experimental effort at JLab \cite{Aznauryan:2011ub} to extract the electromagnetic form factors
and helicity amplitudes of baryonic resonances in the regime where non-perturbative effects are
dominant and perturbative QCD predictions fail.

\medskip

The paper is organized as follows: In section 2 we introduce the current matrix decomposition for resonance production in proton electromagnetic scattering and present our theoretical
results for electromagnetic form factors and helicity amplitudes in the Breit frame. Moreover, we study the contributions of resonance production to the proton structure functions defined
in Deep Inelastic Scattering (DIS). Here, we focus on the production of negative parity baryonic resonances and their contributions, but for completeness,
we also provide a review of our previous results for positive parity baryon resonances.
Section 3 contains a detailed computation of the electromagnetic currents in the holographic Sakai-Sugimoto model and a subsequent calculation of Dirac and Pauli form factors
in the holographic setup. This is done in a unified manner for both positive and negative parity baryon resonances.
In section 4 we present our numerical results for the generalized form factors, helicity amplitudes and proton structure functions for the special case of negative parity baryon resonances and compare them to experimental results.
Section 5 offers some conclusions and an outlook. Appendix A reviews different frames utilized in this article
while appendix B gives technical details on the limits relevant to the model at hand.

\medskip

Previous holographic calculations on electromagnetic form factors of baryons can be found in
\cite{Hong:2007ay,Hashimoto:2008zw,Kim:2008pw,arXiv:0903.4818,arXiv:0904.3710,arXiv:0904.3731}.
DIS structure functions from holography were first obtained in \cite{Polchinski:2002jw}.
Further developments include the large $x$ regime \cite{BallonBayona:2007qr,BallonBayona:2008zi,Pire:2008zf,BallonBayona:2010ae}
as well as the small $x$ regime \cite{Brower:2006ea,Hatta:2007he,BallonBayona:2007rs,arXiv:0804.1562,Cornalba:2009ax,Brower:2010wf};
DIS structure functions have also been calculated for strongly coupled plasmas \cite{Hatta:2007cs,Bayona:2009qe,Iancu:2009py,Bu:2011my}.

\section{Form factors and helicity amplitudes}

\subsection{Dirac and Pauli form factors}

We want to describe the electromagnetic interaction of a spin $1/2$ baryon in the case where, as a result of the interaction, a spin $1/2$ baryonic resonance is produced.
This baryonic transition is described by an electromagnetic current evaluated between the initial and final states. In our approach, we embed the  electromagnetic current
in a vectorial $U(2)$ symmetry present in any effective description of large-$N_c$ QCD with chiral symmetry breaking. Then we define the electromagnetic current as a
linear combination of flavour currents:
\beqa
{\cal J}^\mu = \sum_a c_a J_V^{\mu,a} \qquad c_0 = 1/N_c \, , \, c_3 =1 \, ,  \, c_1 = c_2 = 0 \,. \label{emcurrent}
\eeqa
Now we evaluate the flavour currents $J_V^{\mu,a}$  between the initial and final baryonic states.

\medskip
\noindent
{\bf Positive parity resonances}

\medskip
\noindent
When the final baryonic state has positive parity we can expand the current matrix element as
\beqa
\langle p_{\scst   X} , B_{\scst   X} , s_{\scst   X}  \vert J_V^{\mu,a} (0) \vert p , B, s \rangle &=& \frac{i}{2 (2 \pi)^3}  (\tau^a)_{I_3^{\scst   X} I_3}
\left ( \eta^{\mu \nu}  - \frac{ q^\mu q^\nu}{q^2} \right )
 \bar u (p_{\scst   X}  , s_{\scst   X}) \Big [  \gamma_\nu F^{D,a}_{B B_{\scst   X}}(q^2)  \cr
&+& \kappa_B  \sigma_{\nu \lambda} q^\lambda  F^{P,a}_{B B_{\scst   X}} (q^2)  \Big ] u (p , s)  \, , \label{matrixelement}
\eeqa
where
\beqa
q^\mu &=& (p_{  X} - p)^\mu \quad , \quad \kappa_B = \frac{1}{m_B + m_{B_{  X}} } \, , \cr
(\tau^0)_{I_3^{  X} I_3} &=& \delta_{I_3^{  X} I_3} \quad , \quad
(\tau^a)_{I_3^{  X} I_3} = (\sigma^a)_{I_3^{  X} I_3} \quad a=(1,2,3) \,,
\eeqa
and $\sigma^a$ are the Pauli matrices. Here we are using the metric $\eta^{\mu \nu} = {\rm diag}(-,+,+,+)$ and
we adopt the following convention for spinors and gamma matrices:
\beqa
u(p,s) &=& \frac{1}{\sqrt{2E}} \left( \begin{array}{c} f \chi_s (\vec{p})  \\ \frac{\vec{p} \cdot \vec{\sigma}}{f} \chi_s (\vec{p}) \end{array} \right)
\quad , \quad
u(p_{\scst   X},s_{\scst   X}) = \frac{1}{\sqrt{2 E_{\scst   X}}} \left( \begin{array}{c} f_{\scst   X} \chi_{s_{\scst   X}} (\vec{p}_{\scst   X})
 \\ \frac{\vec{p}_{\scst   X} \cdot \vec{\sigma}}{f_{\scst   X}} \chi_{s_{\scst   X}} (\vec{p}_{\scst   X})  \end{array} \right)
\, , \cr
\gamma^0  &=& - i \left( \begin{array}{cc} 1 & 0  \\ 0 & - 1 \end{array} \right) \quad, \quad \gamma^i  = - i \left( \begin{array}{cc} 0 & \sigma^i  \\ -\sigma^i  & 0 \end{array} \right) \, , \cr
\sigma^{\mu \nu} &=& \frac{i}{2} \left [ \gamma^\mu , \gamma^\nu \right ] \quad , \quad
\gamma_5 = i \gamma^0 \gamma^1 \gamma^2 \gamma^3 = \left( \begin{array}{cc} 0 & 1  \\ 1 & 0  \end{array} \right) \,, \label{GammaMatrices}
\eeqa
where
\beqa
f = \sqrt{E + m_B} \quad , \quad f_{\scst   X} = \sqrt{E_{\scst   X} + m_{B_{\scst   X}}} \, .
\eeqa

\medskip

In this article we are interested in the case where $I_3^{\scst   X} = I_3 =1/2$. In this case the baryonic states satisfy the relation \cite{Bayona:2011xj} :
\beqa
\langle p_{\scst   X}, B_{\scst  X} , S_{\scst  X} \vert  \vec{p} , B , s \rangle
&=&  \delta^3 (\vec {p}_{\scst   X} - \vec{p}) \delta_{s_{\scst  X} s}
\delta_{B_{\scst  X} B} \, .
\eeqa

The spinors  $\chi_s (\vec{p})$ are defined as the eigenstates of the helicity equation of the initial state:
\beqa
\vec{p} \cdot \vec{\sigma}  \chi_s(\vec{p})  = s | \vec{p} | \chi_s(\vec{p}),  \quad s = (+,-) \, , \label{helinitial}
\eeqa
Similary, the  spinors $ \chi_{ s_{X}} (\vec{p}_{X})$ are defined by the helicity equation for the final state:
\beqa
\vec{p}_{\scst  X} \cdot \vec{\sigma} \chi_{ s_{X}} (\vec{p}_{X}) =   s_{\scst  X} | \vec{p}_{\scst  X} |  \chi_{ s_{\scst  X}} (\vec{p}_{\scst  X}) \,. \label{helfinal}
\eeqa

In order to get standard relativistic normalizations
we need to transform the spinors and baryon states as \cite{Bayona:2011xj}
\beqa
u (p , s)  \to \frac{1}{\sqrt{2 E}} \, u (p,s) \quad , \quad \vert  p , B , s \rangle
\to \frac{1}{\sqrt{2 E} (2 \pi)^{3/2}}  \vert  p , B , s \rangle \,. \label{spinortransform}
\eeqa
Using (\ref{emcurrent}), (\ref{matrixelement}) and (\ref{spinortransform}) we obtain for $I_3=I_3^{ X}=1/2$,
\beqa
\langle p_{  X} , B_{  X} , s_{  X}  \vert {\cal J}^\mu (0) \vert p , B, s \rangle &=& i
\left ( \eta^{\mu \nu}  - \frac{ q^\mu q^\nu}{q^2} \right )
 \bar u (p_{  X}  , s_{  X}) \Big [  \gamma_\nu F^{D}_{B B_{  X}}(q^2)  \cr
&&  \quad \quad + \kappa_B  \sigma_{\nu \lambda} q^\lambda  F^{P}_{B B_{  X}} (q^2)  \Big ] u (p , s)  \, ,
\label{matrixelement2}
\eeqa
where
\beqa
F^{D}_{B B_{  X}}(q^2) = \frac12 \sum_a c_a F^{D,a}_{B B_{  X}}(q^2) \quad , \quad
F^{P}_{B B_{  X}}(q^2) = \frac12 \sum_a c_a F^{P,a}_{B B_{  X}}(q^2) \,,
\eeqa
are the generalized Dirac and Pauli form factors that describe the production of positive parity baryons.

\medskip
\noindent
{\bf Negative parity resonances}

\medskip
\noindent
A good expansion for the flavor current matrix element  in the case when the final baryonic state has negative parity is given by
\beqa
_5 \langle p_{\scst   X} , B_{\scst   X} , s_{\scst   X}  \vert J_V^{\mu,a} (0) \vert p , B, s \rangle &=& \frac{i}{2 (2 \pi)^3}  (\tau^a)_{I_3^{\scst   X} I_3}
\left ( \eta^{\mu \nu}  - \frac{ q^\mu q^\nu}{q^2} \right )
 \bar u (p_{\scst   X}  , s_{\scst   X}) \Big [  \gamma_\nu \tilde F^{D,a}_{B B_{\scst   X}}(q^2)  \cr
&+& \kappa_B  \sigma_{\nu \lambda} q^\lambda  \tilde F^{P,a}_{B B_{\scst   X}} (q^2)  \Big ] \gamma_5 u (p , s)  \, . \label{matrixelementgamma5}
\eeqa
Alternatively, if we want to associate the chirality matrix $\gamma_5$ with the final state (which is the non-trivial state) we can write the current matrix element as
\beqa
_5 \langle p_{\scst   X} , B_{\scst   X} , s_{\scst   X}  \vert J_V^{\mu,a} (0) \vert p , B, s \rangle &=& \frac{i}{2 (2 \pi)^3}  (\tau^a)_{I_3^{\scst   X} I_3}
\left ( \eta^{\mu \nu}  - \frac{ q^\mu q^\nu}{q^2} \right )
 \bar u (p_{\scst   X}  , s_{\scst   X}) \gamma_5 \cr
&& \times \Big[  - \gamma_\nu \tilde F^{D,a}_{B B_{\scst   X}}(q^2)
+ \kappa_B  \sigma_{\nu \lambda} q^\lambda  \tilde F^{P,a}_{B B_{\scst   X}} (q^2)  \Big ]  u (p , s)  \, .
\eeqa
Transforming the spinors and states as (\ref{spinortransform}), we get for $I_3=I_3^{ X}=1/2$,
\beqa
_5 \langle p_{  X} , B_{  X} , s_{  X}  \vert {\cal J}^\mu (0) \vert p , B, s \rangle &=& i
\left ( \eta^{\mu \nu}  - \frac{ q^\mu q^\nu}{q^2} \right )
 \bar u (p_{  X}  , s_{  X}) \Big [  \gamma_\nu \tilde F^{D}_{B B_{  X}}(q^2)  \cr
&&  \quad \quad + \kappa_B  \sigma_{\nu \lambda} q^\lambda  \tilde F^{P}_{B B_{  X}} (q^2)  \Big ]
\gamma_5 u (p , s)  \, , \label{matrixelement2gamma5}
\eeqa
where
\beqa
\tilde F^{D}_{B B_{  X}}(q^2) = \frac12 \sum_a c_a \tilde F^{D,a}_{B B_{  X}}(q^2) \quad , \quad
\tilde F^{P}_{B B_{  X}}(q^2) = \frac12 \sum_a c_a \tilde F^{P,a}_{B B_{  X}}(q^2) \,,
\eeqa
are the generalized Dirac and Pauli form factors that describe the production of negative parity baryons.

\subsection{The Breit frame}

It is usually convenient to work in the Breit frame where
\beqa
p^\mu = (E , 0 , 0 , p ) \quad , \quad
q^\mu = (0 , 0 , 0 , - 2 x p ) \quad , \quad p^\mu_{\scst  X} = (E , 0 ,0 , p (1 - 2 x))  \,.
\eeqa
The details of this frame are given in the Appendix. In the Breit frame, we obtain the following helicity equation:
\beqa
\vec{p}_{\scst  X} \cdot \vec{\sigma}  \chi_{ s_{\scst  X}} (\vec{p}) &=& (1 - 2 x) p \sigma^3 \chi_{ s_{\scst  X}}(\vec{p}) \cr
&=&  s_{\scst  X} (1 - 2 x) | \vec{p} | \chi_{ s_{\scst  X}} (\vec{p}) \, ,
\eeqa
Using the relation $| \vec{p}_{\scst  X} |= | 1 - 2 x || \vec{p} |$ and (\ref{helfinal}), we identify two situations:
\beqa
{\rm If} \, \, 1 - 2 x > 0   &\to& \chi_{ s_{\scst  X}} (\vec{p}_{\scst  X})  = \chi_{ s_{\scst  X}} (\vec{p}) \, , \cr
{\rm If} \, \,  1 - 2 x < 0   &\to& \chi_{ s_{\scst  X}} (\vec{p}_{\scst  X}) = \chi_{ - s_{\scst  X}} (\vec{p}) \, . \label{split}
\eeqa

\medskip
\noindent
{\bf Positive parity resonances}

\medskip
\noindent
Using the helicity equations we can calculate the current matrix elements  in the Breit frame (for details see \cite{Bayona:2011xj}).
The result is
\beqa
\langle p_{X} , B_{X} , s_{X}  \vert J_V^{0,a} (0) \vert p , B, s \rangle
&=&  \frac{1}{2 (2 \pi)^3} (\tau^a)_{I_3^{\scst   X} I_3}  \chi^\dag_{s_{X}} (\vec{p}_{X}) \chi_s (\vec{p})  \cr
&\times&  \left [ \alpha F^{D,a}_{B B_{\scst   X}}(q^2) - \beta q^2 \kappa_B F^{P,a}_{B B_{\scst   X}}(q^2) \right ]
\, ,    \label{finalcurrent0}
\eeqa
\beqa
\langle p_{\scst   X} , B_{\scst   X} , s_{\scst   X}  \vert J_V^{i,a} (0) \vert p , B, s \rangle
&=&  - \frac{i}{2 (2 \pi)^3} (\tau^a)_{I_3^{\scst   X} I_3} \epsilon^{ijk} q_j
\chi^\dag_{s_{X}} (\vec{p}_{X})  \sigma_k \chi_s(\vec{p})  \cr
&\times& \Big [ \beta F^{D,a}_{B B_{\scst   X}}(q^2) + \alpha \kappa_B F^{P,a}_{B B_{\scst   X}}(q^2) \Big ]
\, , \label{finalcurrenti}
\eeqa
where
\beqa
\alpha &=& \left (\frac{1}{2E} \right ) \left ( \frac{f}{f_{\scst  X}}\right ) \left [ f_{\scst  X}^2  + (1-2x) \frac{|\vec{p}|^2}{f^2} \right ] \, , \label{defalpha} \\
\beta &=&  \left (\frac{1}{2E} \right ) \left ( \frac{f}{f_{\scst  X}}\right ) \left ( \frac{1}{2x} \right ) \left [ \frac{f_{\scst  X}^2 }{f^2} + 2x - 1 \right ] \label{defbeta} \,.
\eeqa

\medskip

\noindent
{\bf Negative parity resonances}

\medskip
\noindent
Note that
\beqa
\gamma_5 u (p,s) =  \frac{1}{\sqrt{2E}} \left( \begin{array}{c}  \frac{ s |\vec{p}|}{f} \chi_s (\vec{p}) \\ f \chi_s (\vec{p}) \end{array} \right)
\,=\,   \frac{1}{\sqrt{2E}} \left( \begin{array}{c} \tilde f \chi_s (\vec{p})  \\ \frac{ s |\vec{p}|}{ \tilde f} \chi_s (\vec{p}) \end{array} \right) \, ,
\eeqa
where
\beqa
 \tilde f := \frac{ s |\vec{p}|}{f} \, .
\eeqa
Therefore, we can recycle the results obtained in the positive parity case by substituting $f$
by $\tilde f$ in all the calculations.  Then it is not difficult to check that in the Breit frame the
current matrix element takes the form
\beqa
_5  \langle p_{\scst   X} , B_{\scst   X} , s_{\scst   X}  \vert J_V^{0,a} (0) \vert p , B, s \rangle
&=&   - \left ( \frac{1}{2 x } \right )\frac{1}{2 (2 \pi)^3}  (\tau^a)_{I_3^{\scst   X} I_3} q^i \chi^\dag_{s_{\scst  X}}(\vec{p}_{\scst  X})  \sigma_i \chi_s  (\vec{p}) \cr
&\times& \Big  [  \hat \alpha \tilde F^{D,a}_{B B_{\scst   X}}(q^2) -   \hat \beta \, q^2  \kappa_B \tilde F^{P,a}_{B B_{\scst   X}}(q^2) \Big ] \, , \label{chfinalcurrent0}
\eeqa
\beqa
_5 \langle p_{\scst   X} , B_{\scst   X} , s_{\scst   X}  \vert J_V^{i,a} (0) \vert p , B, s \rangle
&=& \left ( \frac{1}{2 x} \right )  \frac{\vec{q}^2 }{2 (2 \pi)^3} (\tau^a)_{I_3^{\scst   X} I_3} \left ( \delta^{ij} - \frac{q^i q^j}{\vec{q}^2} \right ) \chi^\dag_{s_{\scst  X}} (\vec{p}_{\scst  X})  \sigma_j \chi_s (\vec{p}) \cr
&\times& \Big [  \hat \beta \tilde F^{D,a}_{B B_{\scst   X}}(q^2)
+  \hat \alpha \kappa_B \tilde F^{P,a}_{B B_{\scst   X}}(q^2) \Big ] \, .
 \label{chfinalcurrenti}
\eeqa
where
\beqa
\hat \alpha &:=& \left ( \frac{f }{ f_{\scst  X}  } \right ) \left (\frac{1}{2E} \right )  \left [ \frac{f_{\scst  X}^2}{f^2}  +  1 - 2x \right ]
 \, , \label{hatalpha} \\
\hat \beta &:=&  \frac{ 1 }{|\vec{p}|^2} \left ( \frac{f}{ f_{\scst  X}} \right ) \left (\frac{1}{2E} \right ) \left ( \frac{1}{2x} \right ) \left [ f_{\scst  X}^2  - (1 - 2x) \frac{|\vec{p}|^2}{f^2}  \right ]
 \,. \label{hatbeta}
\eeqa

\subsection{Helicity amplitudes}

\medskip

In order to establish a simple connection between the Dirac and Pauli form factors and the more commonly used
helicity amplitudes, we first need to review some Gordon identities.
We start with a generalized Gordon identity
\beqa
p^{\scst  X}_\nu \gamma^\nu \gamma^\mu + p_\nu \gamma^\mu \gamma^\nu &=& p^{\scst  X}_\nu \left  ( \frac12 \{ \gamma^\nu , \gamma^\mu \} + \frac12 [ \gamma^\nu , \gamma^\mu ] \right ) +
p_\nu \left  ( \frac12 \{ \gamma^\mu , \gamma^\nu \} + \frac12 [ \gamma^\mu , \gamma^\nu ] \right ) \cr
&=& p^{\scst  X}_\nu \left ( \eta^{\mu \nu} + i \sigma^{\mu \nu} \right ) + p_\nu \left ( \eta^{\mu \nu} - i \sigma^{\mu \nu} \right ) \cr
&=& (p_{\scst  X} + p)^{\mu} + i \sigma^{\mu \nu} q_\nu \, . \label{GenGordon}
\eeqa
Evaluating (\ref{GenGordon}) on the initial and final spinor and using  the Dirac equation we get the Gordon decomposition for positive parity resonances:
\beqa
\bar u(p_{\scst  X}, s_{\scst  X}) \gamma^\mu u(p,s) =
 - \frac{i}{m_{B_{\scst  X}} + m_B} \bar u(p_{\scst  X}, s_{\scst  X})
\left [ (p_{\scst  X} + p)^\mu + i \sigma^{\mu \nu} q_\nu \right ] u(p,s) \, , \label{Gordonidpos}
\eeqa
On the other hand, if we multiply (\ref{GenGordon}) by $\gamma_5$ on the right, evaluate it on the initial and final spinors, and finally use the Dirac equation, we get
the Gordon decomposition for the negative parity case,
\beqa
\bar u(p_{\scst  X}, s_{\scst  X}) \gamma^\mu \gamma_5 u (p,s) =
 - \frac{i}{ m_{B_{\scst  X}} - m_B} \bar u(p_{\scst  X}, s_{\scst  X})
\left [ (p_{\scst  X} + p)^{\mu} + i \sigma^{\mu \nu} q_\nu \right ] \gamma_5 u(p,s) \,. \label{Gordonidneg}
\eeqa

\medskip
\noindent
{\bf Positive parity resonances}

\medskip
\noindent
First we  define the $G_1(q^2)$ and $G_2(q^2)$ form factors through the vector current decomposition \cite{Aznauryan:2008us}
\beqa
\langle p_{\scst   X} , B_{\scst   X} , s_{\scst   X}  \vert {\cal J}^{\mu} (0) \vert p , B, s \rangle &\,=\,& i  \, \bar u(p_{\scst  X}, s_{\scst  X})
\Big \{ \left [ \eta^{\mu \nu} - \frac{q^\mu q^\nu}{q^2} \right ] \gamma_\nu q^2 G_1 (q^2) \cr
&&+ \frac12 \left [ (p_{\scst  X}^2 - p^2) \gamma^\mu - q_\nu \gamma^\nu (p_{\scst  X} + p)^\mu  \right ] G_2(q^2) \Big \} u (p,s) \, . \cr && \label{Azncurrent}
\eeqa
Using the Gordon identity (\ref{Gordonidpos}) and the Dirac equation we can rewrite  the current as in  (\ref{matrixelement2}). This way we get the Dirac and Pauli form factors
in terms of the $G_1(q^2)$ and $G_2(q^2)$ form factors:
\beqa
F^D_{B B_{\scst  X}}(q^2) &=& q^2 G_1(q^2) \cr
F^P_{B B_{\scst  X}}(q^2) &=& - \frac12 (m_{B_{\scst  X}}^2 - m_B^2) G_2(q^2) \, .
\eeqa
According to \cite{Aznauryan:2008us}, the transverse helicity amplitude ${\cal A}_{1/2}(q^2)$ is defined by
\beqa
{\cal A}_{1/2}(q^2) &=& \sqrt{ \frac{E_R - m_B}{2 m_B K}} \left [ q^2 G_1(q^2) - \frac12 (m_{B_{\scst  X}}^2 - m_B^2) G_2(q^2) \right ] \cr
&=&  \sqrt{ \frac{E_R - m_B}{2 m_B K}} \left [ F^D_{B B_{\scst  X}}(q^2) + F^P_{B B_{\scst  X}}(q^2) \right ] \, , \label{A12Az}
\eeqa
where
\beqa
K = \frac{m_{B_{\scst  X}}^2 - m_B^2}{2 m_{B_{\scst  X}}} \, , \label{Kdef}
\eeqa
and $E_R$ is the proton energy in the resonance rest frame.
Details of the resonance rest frame are given in appendix A.2.

The helicity amplitude ${\cal A}_{1/2}(q^2)$ can be rewritten as \cite{Carlson:1998gf}
\beqa
{\cal A}_{1/2}(q^2) = \sqrt{ \frac{m_B }{m_{B_{\scst  X}}^2 - m_B^2 }} G^+_{B B_{\scst  X}}(q^2) \, , \label{A12Carl}
\eeqa
where
\beqa
G^+_{B B_{\scst  X}}(q^2) = \frac{\zeta}{m_B}  \left [ F^D_{B B_{\scst  X}}(q^2) + F^P_{B B_{\scst  X}}(q^2) \right ] \, , \label{G+}
\eeqa
and
\beqa
\zeta := \sqrt{ m_{B_{\scst  X}} (E_R - m_B)} = \frac{1}{\sqrt{2}} \left [ (m_{B_{\scst  X}} - m_B)^2 + q^2 \right ]^{1/2} \,. \label{zeta}
\eeqa

The longitudinal helicity amplitude ${\cal S}_{1/2}(q^2)$ is given by  \cite{Aznauryan:2008us},
\beqa
{\cal S}_{1/2}(q^2) &=&  \sqrt{\frac{E_R - m_B}{ m_B K} } \frac{ | \vec{q}_R |}{2} \left [ (m_{B_X} + m_B) G_1(q^2) + \frac12 (m_{B_X}- m_B) G_2 (q^2) \right ] \cr
&=&  \sqrt{\frac{E_R - m_B}{ m_B K} } \frac{ | \vec{q}_R |}{2} \left [ \frac{m_{B_X} + m_B}{q^2} F^D_{B B_{\scst  X}}(q^2)  - \frac{1}{m_{B_X} + m_B}  F^P_{B B_{\scst  X}}(q^2)  \right ] \, ,
\eeqa
where $\vec{q}_R$ is the spatial momentum of the virtual photon in the resonance rest frame.
According to \cite{Stoler:1993yk}, this amplitude can be rewritten as
\beqa
{\cal S}_{1/2}(q^2) =  \sqrt{\frac{m_B}{m_{B_X}^2 - m_B^2}} \frac{ |\vec{q}_R|}{\sqrt{q^2}} G^0_{B B_{\scst  X}}(q^2) \,, \label{relS12G0}
\eeqa
where
\beqa
G^0_{B B_{\scst  X}}(q^2) = \sqrt{\frac{q^2}{2}} \frac{\zeta}{m_B} \left [ \frac{m_{B_X} + m_B}{q^2} F^D_{B B_{\scst  X}}(q^2)  - \frac{1}{m_{B_X} + m_B}  F^P_{B B_{\scst  X}}(q^2)  \right ] \, .
\label{G0}
\eeqa

\medskip
\noindent
{\bf Negative parity resonances}

\medskip
\noindent
In analogy with the previous case we  define the $\tilde G_1(q^2)$ and $\tilde G_2(q^2)$ negative parity form factors through the
vector current decomposition as in \cite{Aznauryan:2008us},
\beqa
{}_5 \langle  p_{\scst   X} , B_{\scst   X} , s_{\scst   X} \vert {\cal J}^{\mu} (0) \vert p , B, s \rangle &\,=\,& - i  \, \bar u(p_{\scst  X}, s_{\scst  X})
\Big \{ \left ( \eta^{\mu \nu} - \frac{q^\mu q^\nu}{q^2} \right ) \gamma_\nu q^2 \tilde G_1 (q^2) \cr
&&+ \frac12 \left [ (p_{\scst  X}^2 - p^2) \gamma^\mu - q_\nu \gamma^\nu (p_{\scst  X} + p)^\mu  \right ] \tilde G_2(q^2) \Big \} \gamma_5 u (p,s) \, . \cr && \label{Azncurrentgamma5}
\eeqa
Using the Gordon identity (\ref{Gordonidneg}) and the Dirac equation we can rewrite (\ref{Azncurrentgamma5}) as in (\ref{matrixelement2gamma5}). Therefore we obtain the relations
\beqa
\tilde F^D_{B B_{\scst  X}}(q^2) &=& - q^2 \tilde G_1(q^2) \cr
\tilde F^P_{B B_{\scst  X}}(q^2) &=&  \frac12 (m_{B_{\scst  X}} + m_B)^2 \tilde G_2(q^2) \, .
\eeqa
Now let us write the expressions for the helicity amplitudes. According to \cite{Aznauryan:2008us}, the helicity amplitudes ${\cal A}_{1/2}(q^2)$ are given by
\beqa
\tilde {\cal A}_{1/2}(q^2) &=&  \sqrt{ \frac{E_R + m_B}{2 m_B K}} \left [ q^2 \tilde G_1(q^2) - \frac12 (m_{B_{\scst  X}}^2 - m_B^2) \tilde G_2(q^2) \right ] \cr
&=& -  \sqrt{ \frac{E_R + m_B}{2 m_B K}} \left [ \tilde F^D_{B B_{\scst  X}}(q^2) + \frac{m_{B_{\scst  X}} - m_B}{m_{B_{\scst  X}} + m_B} \tilde F^P_{B B_{\scst  X}}(q^2) \right ] \, , \label{A12Azgamma5}
\eeqa
where $K$ is given by (\ref{Kdef}) and $E_R$ is the proton energy in the resonance rest frame (defined in appendix A.2).

Using the analog of (\ref{A12Carl}), we get
\beqa
\tilde G^+_{B B_{\scst  X}}(q^2) = - \frac{\tilde \zeta }{m_B} \left [ \tilde F^D_{B B_{\scst  X}}(q^2) + \frac{m_{B_{\scst  X}} - m_B}{m_{B_{\scst  X}} + m_B} \tilde F^P_{B B_{\scst  X}}(q^2) \right ] \, ,
\label{tildeG+}
\eeqa
where
\beqa
\tilde \zeta := \sqrt{m_{B_{\scst  X}} (E_R + m_B) } &=& \frac{1}{\sqrt{2}} \left [ (m_{B_{\scst  X}} + m_B)^2 + q^2 \right ]^{1/2}  \, . \label{tildezeta}
\eeqa

The helicity amplitude $\tilde {\cal S}_{1/2}(q^2)$ is given by \cite{Aznauryan:2008us}
\beqa
\tilde {\cal S}_{1/2}(q^2) &=& -  \sqrt{\frac{E_R + m_B}{ m_B K} } \frac{ | \vec{q}_R |}{2} \left [ (m_{B_X} - m_B) \tilde G_1(q^2) + \frac12 (m_{B_X} + m_B) \tilde G_2 (q^2) \right ] \cr
&=&   \sqrt{\frac{E_R + m_B}{ m_B K} } \frac{ | \vec{q}_R |}{2} \left [ \frac{m_{B_X} - m_B}{q^2} \tilde F^D_{B B_{\scst  X}}(q^2)  - \frac{1}{m_{B_X} + m_B}  \tilde F^P_{B B_{\scst  X}}(q^2)  \right ] \, ,
\eeqa
where $\vec{q}_R$ is the spatial momentum of the virtual photon in the resonance rest frame. Using the negative parity analog of (\ref{relS12G0})  we get
\beqa
\tilde G^0_{B B_{\scst  X}}(q^2) = \sqrt{\frac{q^2}{2}} \frac{\tilde \zeta}{m_B} \left [ \frac{m_{B_X} - m_B}{q^2} \tilde F^D_{B B_{\scst  X}}(q^2)  - \frac{1}{m_{B_X} + m_B}  \tilde F^P_{B B_{\scst  X}}(q^2)  \right ] \, .
\label{tildeG0}
\eeqa

\subsection{The proton structure functions}

A typical deep inelastic scattering (DIS) process a lepton scatters a hadron via the exchange of a virtual photon. The corresponding
differential cross section is determined by the hadronic tensor,
\begin{equation}\label{eq:Wmunu}
W^{\mu\nu} \, = \, \frac{1}{8 \pi} \sum_s \int d^4 x \, e^{iq\cdot x} \langle p , s \vert \, \Big[ {\cal J}^\mu (x) , {\cal J}^\nu (0) \Big]
\, \vert p , s \rangle,
\end{equation}
where ${\cal J}^\mu(x)$ is the electromagnetic current, $q^\mu$ and $p^\mu$ are the momenta of the  virtual photon and the initial hadron, respectively.

One usually parametrizes DIS using as dynamical variables the Bjorken parameter $x = -\frac{q^2}{2 p \cdot q}$ and the photon virtuality $q^2$. The hadronic tensor can be decomposed in terms of the Lorentz invariant scalar structure functions  ${\cal F}_1 (x,q^2)$ and ${\cal F}_2 (x,q^2)$:
\begin{equation} \label{eq:strfun}
W^{\mu\nu} \, = \, {\cal F}_1 (x,q^2)  \Big( \eta^{\mu\nu} \,-\, \frac{q^\mu q^\nu}{q^2} \, \Big)
\,+\,\frac{2x}{q^2} {\cal F}_2 (x,q^2)  \Big( p^\mu \,+ \, \frac{q^\mu}{2x} \, \Big)
\Big( p^\nu \,+ \, \frac{q^\nu}{2x} \, \Big).
\end{equation}
The standard limit of DIS corresponds to the Bjorken limit of large $q^2$ and fixed $x$.
In this paper we are interested in the regime of small $q^2$ where non-perturbative contributions are relevant (for a review of DIS, see e.g., \cite{Manohar:1992tz}).

The hadronic tensor for  a spin $1/2$ baryon, in the case where one particle is produced in the final state, can be written as
\beqa
W^{\mu \nu} &=& \frac14  \sum_{s,s_{ X}} \sum_{m_{B_X}} \delta \left [ ( p+ q)^2 + m_{B_X}^2  \right ]
\Big [ \langle p , B ,s | {\cal J}^{\mu}(0) | p_X , B_X , s_{ X} \rangle  \langle p_X , B_X , s_{ X} | {\cal J}^{\nu}(0) | p , B , s \rangle  \cr
&& \quad + \langle p , B , s | {\cal J}^{\mu}(0) | p_X , B_X , s_{ X} \rangle_5 \, {}_5 \langle p_X , B_X  , s_{ X} | {\cal J}^{\nu}(0) | p , B \rangle \Big ] \,.
\label{hadronictensor}
\eeqa
Note that we are including the contribution from positive parity resonances as well as negative parity resonances.
 Substituting (\ref{matrixelement2}) and (\ref{matrixelement2gamma5}) into (\ref{hadronictensor}) and using some gamma trace identities  we obtain the proton structure functions \footnote{For more details see \cite{Bayona:2011xj}}
\beqa
{\cal F}_1 (q^2 ,x) &=& F_1 (q^2 ,x) + \tilde F_1 (q^2 ,x) \, , \\
 {\cal F}_2 (q^2 ,x) &=& F_2 (q^2 ,x) + \tilde F_2 (q^2 ,x) \, ,
\eeqa
where
\beqa
F_1 (q^2 ,x) &\,=\,&  \sum_{m_{B_X}} \delta \left [ ( p+ q)^2 + m_{B_X}^2  \right ] m_B^2 ( G^+_{B B_{\scst  X}}(q^2) )^2 \, ,  \\
F_2 (q^2 ,x) &\,=\,&   \sum_{m_{B_X}} \delta \left [ ( p+ q)^2 + m_{B_X}^2  \right ] \left (\frac{q^2}{2x} \right ) \left (1 + \frac{q^2}{4 m_B^2 x^2} \right )^{-1}  \cr
&& \times\left [ (G^+_{B B_{\scst  X}}(q^2))^2 + 2 (G^0_{B B_{\scst  X}}(q^2))^2 \right ] \,,
\eeqa
are the positive parity contributions to the proton structure functions and
\beqa
\tilde F_1 (q^2 ,x) &\,=\,& \sum_{m_{B_X}} \delta \left [ ( p+ q)^2 + m_{B_X}^2  \right ] m_B^2 ( \tilde G^+_{B B_{\scst  X}}(q^2) )^2 \, , \\
\tilde F_2 (q^2 ,x) &\,=\,&  \sum_{m_{B_X}} \delta \left [ ( p+ q)^2 + m_{B_X}^2  \right ] \left (\frac{q^2}{2x} \right ) \left (1 + \frac{q^2}{4 m_B^2 x^2} \right )^{-1}  \cr
&& \quad \times \left [ (\tilde G^+_{B B_{\scst  X}}(q^2))^2 + 2 (\tilde G^0_{B B_{\scst  X}}(q^2))^2 \right ] \,,
\eeqa
are the negative parity contributions to the proton structure functions.

\section{Dirac and Pauli form factors from holography}

\subsection{Review of the Sakai-Sugimoto model}

\subsubsection{D4 -- D8 configuration}

The Sakai-Sugimoto model \cite{Sakai:2004cn,Sakai:2005yt} is the most widely studied string-theoretic model of
large-$N_c$ QCD and has been successfully applied to investigate many of its phenomenological aspects. Its holographic limit describes a stable configuration of $D8-\overline{D8}$ branes embedded into Witten's $D4$ model \cite{Witten:1998qj}. In the following section we will briefly review those features of this model which are important for the investigations carried out in this article.
The geometry of Witten's model is generated by $N_c$ coincident $D4$ branes with a compact spatial direction $\tau$ in type IIA supergravity with the following metric, dilaton
and four-form,
\begin{align}
ds^2 &= \frac{u^{3/2}}{R^{3/2}} \left(\eta_{\mu \nu} dx^{\mu}dx^{\nu} + f(u) d \tau^2 \right) + \frac{R^{3/2}}{u^{3/2}} \frac{du^2}{f(u)}+ R^{3/2}u^{1/2} d\Omega_4^2, \\\nonumber
f(u) &= 1 -  \frac{u_{KK}^3}{u^3}, \quad  e^{\phi} = g_s \frac{u^{3/4}}{R^{3/4}} , \quad F_4  =
\frac{(2 \pi l_s)^3 N_c}{V_{S^4}}\varepsilon_4,\nonumber
\end{align}
where $u_{KK}$ is the radial position of the tip of the cigar geometry generated by the $D4$ branes and $R = \left(\pi g_s N_c \right)^{1/3} \sqrt{\alpha'}$.
To incorporate fundamental (quark and anti-quark) degrees of freedom, one needs to introduce two
 stacks of $N_f$ coincident $D8$ and $\overline{D8}$ flavor branes into the background generated
by the $N_c$ $D4$ branes. The probe condition $N_f  \ll N_c$ ensures that the back reaction of the
flavor branes on the geometry can be safely neglected. It turns out that the solution to the DBI
equations merges the two stacks of $D8$ and $\overline{D8}$ branes in the infrared region (small u),
 resulting in a geometrical realization of chiral symmetry breaking
$U(N_f ) \times U(N_f) \rightarrow U(N_f)$.
The dynamics of the gauge field fluctuations on the $D8/\overline{D8}$ brane embedding is described by the
Dirac-Born-Infeld action, which yields a vector meson effective field theory given by a five
dimensional $U(N_f)$ Yang-Mills-Chern-Simons theory in a curved background. The details of this construction can be found in the original publications by Sakai and Sugimoto
\cite{Sakai:2004cn,Sakai:2005yt}. For the present context, cf. our previous work \cite{Bayona:2011xj}.

\subsubsection{Baryons in the Sakai-Sugimoto model}

Let us describe the ideas behind the construction of holographic baryons.
Recall that, in the confined phase, the Sakai-Sugimoto model reduces to a five-dimensional $U(N_f)$
Yang Mills-Chern Simons (YM-CS) theory.
In this article, we restrict ourselves to the $N_f=2$ case. Then, the $U(2)$ gauge
field $\cA$ can be decomposed as
\begin{eqnarray}\label{eq:Adecom}
\cA=A+\wh A\,\frac{{\bf 1}_2}{2}
=A^i\frac{\tau^i}{2}+\wh A\,\frac{{\bf 1}_2}{2}
=\sum_{a=0}^3 \cA^a\,\frac{\tau^a}{2}\ ,
\end{eqnarray}
where $\tau^i$ ($i=1,2,3$) are Pauli matrices and $\tau^0={\bf 1}_2$
is a unit matrix of dimension 2.
The equations of motion are given by \cite{Hata:2007mb,Hashimoto:2008zw}
\begin{eqnarray}
&&-\kappa \left(
h(z) \partial_{\nu} \wh F^{\mu\nu}+ \partial_z (k(z)\wh F^{\mu z})
\right)
+\frac{N_c}{128\pi^2}\epsilon^{\mu \alpha_2 \ldots \alpha_5}
\left(
F^a_{\alpha_2\alpha_3}F^a_{\alpha_4\alpha_5}
+\wh F_{\alpha_2\alpha_3}\wh F_{\alpha_4\alpha_5}
\right)
=0, \nonumber \\
&& -\kappa\left(
h(z)\nabla_\nu F^{\mu\nu}+\nabla_z (k(z) F^{\mu z})
\right)^a
+\frac{N_c}{64\pi^2}\epsilon^{\mu \alpha_2 \ldots \alpha_5}
F^a_{\alpha_2\alpha_3}\wh F_{\alpha_4\alpha_5}=0, \nonumber \\
&&-\kappa
k(z)\partial_{\nu} \wh F^{z\nu}
+\frac{N_c}{128\pi^2}\epsilon^{z \mu_2 \ldots \mu_5}
\left(
F^a_{\mu_2\mu_3} F^a_{\mu_4\mu_5}
+\wh F_{\mu_2\mu_3} \wh F_{\mu_4\mu_5}
\right)
=0, \nonumber \\
&&-\kappa k(z)
\left(\nabla_\nu F^{z\nu}\right)^a
+\frac{N_c}{64\pi^2} \epsilon^{z \mu_2 \ldots \mu_5}
F^a_{\mu_2\mu_3}\wh F_{\mu_4\mu_5}=0, \label{eqsmotion}
\end{eqnarray}
where $k(z) = 1 + z^2 $ and $h(z) = (1 + z^2)^{-1/3}$ are the 5-d warp factors and $\nabla_\alpha=\partial_\alpha+iA_\alpha$ is the covariant derivative.
The baryon in the Sakai-Sugimoto holographic model is represented by a soliton with nontrivial instanton number in the four-dimensional space parameterized by $x^M$ ($M=1,2,3,z$).
Consequently, the instanton number is interpreted as the baryon number $N_B$, and reads
\begin{eqnarray}
 N_B=\frac{1}{64\pi^2}\int
d^3x dz\, \epsilon_{M_1M_2M_3M_4} F^a_{M_1M_2}F^a_{M_3M_4}
\ .
\label{NB}
\end{eqnarray}
The construction and quantization of solutions to the set of equations (\ref{eqsmotion}) in the large $\lambda$ regime was discussed in great detail in the literature \cite{Hata:2007mb,Hashimoto:2008zw,Bayona:2011xj}. Here we merely state some results that will be important for the purpose of the present work.
The resulting baryon eigenstates are characterized by quantum numbers $B = (l, I_3, n_{\rho}, n_z)$ in addition to their spin s. For example, the baryon wave functions with quantum numbers $B_n = (1,+1/2, 0, n)$ are given by
\begin{equation}
|B_n \uparrow \rangle  \propto R(\rho)\psi_{B_n} (Z)(a_1 + i a_2) ,
\end{equation}
where
\beqa
R(\rho) &=& {\rho}^{-1+2\sqrt{1+N_c^2 /5}}\, e^{-\frac{M_0}{\sqrt{6}}\rho^2},\\\nonumber
\psi_{B_n}(Z) &=& \left(\frac{(2M_0)^{1/4}}{6^{1/8} \pi^{1/4} 2^{n/2} \sqrt{n!}} \right)  H_n \left( \sqrt{2 M_0}6^{-1/4} Z \right)\, e^{-\frac{M_0}{\sqrt{6}}Z^2}.
\eeqa
The mass formula for the baryonic eigenstates (obtained from the quantized Hamiltonian of the system) reads
\begin{equation}
M = M_0 +\sqrt{\frac{(\ell +1)^2}{6}+\frac{2}{15}N_c^2} + \frac{2(n_{\rho}+n_z)+2}{\sqrt{6}}
=: \tilde M_0 + \frac{2 n_z}{\sqrt{6}} \,. \label{baryonmass}
\end{equation}

\subsection{Electromagnetic currents in the Sakai-Sugimoto model}

The holographic currents in the Sakai-Sugimoto model, denoted here by $J_{V (SS)}^{\mu,a}$,
can be obtained using the holographic relations \cite{Hashimoto:2008zw} :
\beqa
J_{V (SS)}^{\mu,a} = - \kappa \Big \{ \lim_{z \to \infty} \left [ k(z) {\cal F}^{\text{cl}}_{\mu z} \right ]
 + \lim_{z \to - \infty} \left [ k(z) {\cal F}^{\text{cl}}_{\mu z} \right ] \Big \} \,, \label{holcurrent}
\eeqa
where ${\cal F}^{\text{cl}}_{\mu z}$ is the field strength associated with the classical field, cf. (2.80-2.85) in \cite{Hashimoto:2008zw}.

We define the baryon states as
\beqa
\vert \vec{p}, B,   s  , I_3 \rangle &=&  \frac{1}{(2 \pi)^{3/2}} e^{ i \vec{p} \cdot \vec{X}} \vert n_B \rangle \vert n_\rho \rangle  \vert  s , I_3 \rangle_R \, , \cr
\vert \vec{p}_{ X}, B_{ X} , s_{ X} , I_3^{ X}  \rangle  &=&  \frac{1}{(2 \pi)^{3/2}} e^{ i \vec{p}_{ X} \cdot \vec{X}} \vert n_{B_{ X}} \rangle \vert n_\rho \rangle
\vert  s_{ X} , I_3^{ X}  \rangle_R \, .
\eeqa
Here we make use of the results and definitions of a recent publication \cite{BoschiFilho:2011hn}, in which a relativistic generalization of baryon states and wave functions was discussed in detail.
In particular, the spin and isospin part was defined as
\beqa
\vert  s , I_3 \rangle_{ R} &=&  \frac{1}{\sqrt{2E}} \left( \begin{array}{c} f \, \vert  s , I_3 \rangle \\ \frac{ s |\vec{p}| }{f} \, \vert  s , I_3 \rangle \end{array} \right) \, , \cr
\langle  s_{ X} , I_3^{ X}  \vert_{ R} &=&  \frac{1}{\sqrt{2 E_{ X}}} \left(  f_{ X} \, \langle  s_{ X} , I^{ X}_3 \vert \quad  - \frac{ s_{ X}|\vec{p}_{ X}| }{f_{ X}}
\, \langle  s_{ X} , I^{ X}_3 \vert  \right) \, ,
\eeqa
where $\vert  s , I_3 \rangle$ and $\langle  s_{ X} , I^{ X}_3 \vert$ are the non-relativistic initial and final states associated with the spin and isospin operators. From this and (\ref{holcurrent}),  one gets \cite{Bayona:2011xj} :
\beqa
\langle J^{0,0}_{V (SS)}(0) \rangle =  \frac{1}{(2 \pi)^3} \frac{N_c}{2}  \langle s_{ X} , I_3^{ X}  \vert s , I_3 \rangle_{ R}   F^1_{B B_{ X}}(\vec{q}^2) \, ,
\eeqa
\beqa
\langle J^{i,0}_{V (SS)}(0) \rangle &=& \frac{1}{(2 \pi)^3} \frac{N_c}{2}  \langle s_{ X} , I_3^{ X} \vert_{ R} \Big \{
F^1_{B B_{ X}}(\vec{q}^2) \left[ \frac{p^i}{M_0}   - \frac{i}{16 \pi^2 \kappa} \epsilon^{ija} q_j S_a \right ]  \cr
&+& \frac{q^i}{M_0} F^3_{B B_{ X}}(\vec{q}^2)   - \frac{1}{16 \pi^2 \kappa } F^2_{B B_{ X}} (\vec{q}^2)(q^i q^a - \vec{q}^2 \delta^{ia} ) S_a  \Big \} \vert s , I_3 \rangle_{ R} \, ,
\eeqa
\beqa
\langle J^{0,c}_{V (SS)}(0) \rangle  &=& 2 \pi^2 \kappa \,  \frac{1}{(2 \pi)^3} \langle  n_\rho \vert \langle  s_{ X} , I_3^{ X}  \vert_{ R} \Big \{
F^1_{B B_{ X}}(\vec{q}^2) \left [ \frac{ I^c    }{2 \pi^2 \kappa}
+ \frac{i}{M_0} \epsilon^{ija} p_i q_j  \rho^2 \tr (\tau^c  {\bf a} \tau_a {\bf a}^{-1} )   \right ]  \cr
&+& F^2_{B B_{ X}}(\vec{q}^2) \Big [ - i q_i  \rho^2 \tr [ \tau^c \partial_0 (  {\bf a} \tau^i {\bf a}^{-1}  )  ] \cr
&+& \frac{1}{M_0} (\vec{P} \cdot \vec{q} q_i - \vec{q}^2 P_i )
 \rho^2 \tr [ \tau^c  {\bf a} \tau^i {\bf a}^{-1}  ]  \Big ]  \Big \} \vert n_\rho \rangle \vert  s , I_3  \rangle_{ R} \, ,
\eeqa
\beqa
\langle J^{i,c}_{V (SS)}(0) \rangle &=&  2 \pi^2 \kappa \, \frac{1}{(2 \pi)^3} \Big [
 i F^1_{B B_{ X}}(\vec{q}^2) \epsilon^{ija} q_j  + F^2_{B B_{ X}}(\vec{q}^2) (q^i q^a - \vec{q}^2 \delta^{ia} ) \Big ] \cr
&\times& \langle n_\rho \vert \rho^2 \vert n_\rho \rangle \langle  s_{ X} , I_3^{ X}  \vert_{ R}  \tr ( \tau^c  {\bf a} \tau_a {\bf a}^{-1} ) \vert  s , I_3  \rangle_{ R} \,.
\eeqa
where
\beqa
F^1_{B B_{ X}}(\vec{q}^2) &=& \sum_n \frac{g_{v^n} \langle n_{B_{ X}} \vert \psi_{2n-1}(Z) \vert n_B \rangle}{\vec{q}^2 + \lambda_{2n-1}}  \cr
F^2_{B B_{ X}}(\vec{q}^2) &=& \sum_n \frac{g_{v^n} \langle n_{B_{ X}} \vert \partial_Z \psi_{2n-1}(Z)  \vert n_B \rangle }{\lambda_{2n-1} (\vec{q}^2 + \lambda_{2n-1})}  \cr
F^3_{B B_{ X}}(\vec{q}^2) &=&\sum_n \frac{g_{v^n} \langle n_{B_{ X}} \vert \partial_Z \psi_{2n-1}(Z) \partial_Z  \vert n_B \rangle }{\lambda_{2n-1} (\vec{q}^2 + \lambda_{2n-1})} \,,
\label{masterformfactors}
\eeqa
and
\beqa
S^a = - i 4 \pi^2 \kappa \rho^2 \tr ( \tau^a {\bf a}^{-1} \dot {\bf a}) \quad , \quad
I^a =  - i 4 \pi^2 \kappa \rho^2 \tr ( \tau^a {\bf a} \dot {\bf a}^{-1} ) \, ,
\eeqa
are the spin and isospin operators. The operators ${\bf a} = a_4 + i a_a \tau^a$ represent the $SU(2)$ orientations of the instanton and $\rho$ is the instanton size. The momentum $\vec{q}$ is the photon momentum defined by $\vec{q} = \vec{p}_{ X} - \vec{p}$.

In order to calculate the expectation values of the holographic currents we need the following identities:
\beqa
 \langle s_{ X} , I_3^{ X}  \vert s , I_3 \rangle_{ R} &=&
\frac{1}{2\sqrt{E_XE}} (ff_X-\tfrac{ss_{ X}|\vec{p}||\vec{p}_{  X}|}{ff_X})\,  \delta_{I_3^{ X} I_3} \chi^\dagger_{s_{ X}}(\vec{p_{ X}})\, \chi_s (\vec{p})\, , \cr
  \langle s_{ X} , I_3^{ X} \vert_{ R} \tr  ( \tau^c  {\bf a} \tau^a {\bf a}^{-1} ) \vert s , I_3 \rangle_{ R} &=&
-\frac{1}{3\sqrt{E_XE}} (ff_X-\tfrac{ss_{ X}|\vec{p}||\vec{p}_{  X}|}{ff_X})\, \tau^c_{I_3^X I_3} \, \chi^\dagger_{s_{ X}}(\vec{p_{ X}})\,\sigma^a \chi_s (\vec{p})\, , \cr
\langle s_{ X} , I_3^{ X}  \vert_{ R} \, I^c \, \vert  s , I_3 \rangle_{ R} &=&
\frac{1}{4 \sqrt{E_X E}} (ff_X-\tfrac{ss_{ X}|\vec{p}||\vec{p}_{  X}|}{ff_X})\,  (\tau^c)_{I_3^{ X} I_3} \chi^\dagger_{s_{ X}}(\vec{p_{ X}})\, \chi_s (\vec{p})\, , \cr
\langle s_{ X} , I_3^{ X}  \vert_{ R} \, S_a \, \vert  s , I_3 \rangle_{ R} &=&
\frac{1}{4 \sqrt{E_X E}} (ff_X-\tfrac{ss_{ X}|\vec{p}||\vec{p}_{  X}|}{ff_X})\,  \delta_{I_3^{ X} I_3} \chi^\dagger_{s_{ X}}(\vec{p_{ X}})\, \sigma_a \chi_s (\vec{p})\, , \cr
\langle s_{ X} , I_3^{ X} \vert_{ R} \tr  ( \tau^3  \partial_0 ( {\bf a} \tau^i {\bf a}^{-1}) ) \vert s , I_3 \rangle_{ R} &=&
\frac{i}{ M_0 \rho^2 \sqrt{E_XE}} (ff_X-\tfrac{ss_{ X}|\vec{p}||\vec{p}_{  X}|}{ff_X}) \cr
&\times& (\tau^3)_{I_3^{ X} I_3} \,  \chi^\dagger_{s_{ X}}(\vec{p_{ X}})\,\sigma^i \chi_s (\vec{p}) \, . \label{usefulid}
\eeqa
The last identity can be obtained by  first noticing that
\beqa
\tr  ( \tau^c  \partial_0 ( {\bf a} \tau^i {\bf a}^{-1}) ) &=& - \frac{2 i }{M_0 \rho^2}
\Big \{  \left (a_4 \frac{\partial}{\partial a_4} - a_a \frac{\partial}{\partial a_a}  \right ) \delta^{ic}
+ a_i \frac{ \partial}{ \partial a_c} + a_c \frac{ \partial}{ \partial a_i} \cr
&-& \epsilon^{ica} \left (a_a \frac{\partial}{\partial a_4} - a_4 \frac{\partial}{\partial a_a}  \right ) \Big \} \,.
\eeqa

Using the identities (\ref{usefulid}), we get in the Breit frame
\beqa
\langle J^{0,0}_{V (SS)}(0) \rangle &=&  \frac{N_c}{2 (2 \pi)^3} \xi  \delta_{I_3^{ X} I} \chi^\dagger_{s_{ X}}(\vec{p_{ X}}) \chi_s (\vec{p})  F^1_{B B_{ X}}(\vec{q}^2) \, , \cr
\langle J^{i,0}_{V (SS)}(0) \rangle &=& \frac{N_c}{2 (2 \pi)^3 M_0}   \delta_{I_3^{ X} I} \chi^\dagger_{s_{ X}}(\vec{p_{ X}}) \Big \{
 q^i \left [ F^3_{B B_{ X}}(\vec{q}^2) - \frac{1}{2x} F^1_{B B_{ X}}(\vec{q}^2) \right ] \xi  \cr
&-& \frac{i}{4} \alpha \epsilon^{ija} q_j \sigma_a F^1_{B B_{ X}}(\vec{q}^2)  \Big \} \chi_s(\vec{p}) \, , \cr
\langle J^{0,c}_{V (SS)}(0) \rangle  &=& \frac{\xi}{2 (2 \pi)^3} ( \tau^c )_{I_3^{ X} I_3}  \chi^\dagger_{s_{ X}}(\vec{p_{ X}}) \chi_s (\vec{p})  F^1_{B B_{ X}}(\vec{q}^2)  \, , \cr
\langle J^{i,c}_{V (SS)}(0) \rangle &=&  - i \frac{\alpha}{2 (2 \pi)^3} \left ( \frac{M_0}{3} \right )  \langle n_\rho \vert \rho^2 \vert n_\rho \rangle \,
   (\tau^c)_{I_3^{ X} I_3} \cr
&\times& \epsilon^{ija} q_j  \chi^\dagger_{s_{ X}}(\vec{p_{ X}}) \sigma_a  \chi_s (\vec{p})  F^1_{B B_{ X}}(\vec{q}^2) \, , \, \label{SSCurrentBreitPosParity}
\eeqa
when the final state is a positive parity resonance, and
\beqa
_5 \langle J^{0,0}_{V (SS)}(0) \rangle &=& 0 \, , \cr
_5 \langle J^{i,0}_{V (SS)}(0) \rangle &=& \frac{N_c }{8 (2 \pi)^3 M_0} \vec{q}^2 \left ( \delta^{ia} - \frac{q^i q^a}{\vec{q}^2} \right )
\delta_{I_3^{ X} I} \chi^\dagger_{s_{ X}}(\vec{p_{ X}})\, \sigma_a \chi_s (\vec{p})  \, \alpha  F^2_{B B_{ X}}(\vec{q}^2) \, ,  \cr
_5 \langle J^{0,3}_{V (SS)}(0) \rangle  &=& \frac{1}{2 (2 \pi)^3}  (\tau^3)_{I_3^{ X} I_3} \, q_i \chi^\dagger_{s_{ X}}(\vec{p_{ X}})\, \sigma_i \chi_s (\vec{p}) \, \xi F^2_{B B_{ X}}(\vec{q}^2) , \cr
_5 \langle J^{i,3}_{V (SS)}(0) \rangle &=&  \frac{1}{2 (2 \pi)^3} (\tau^3)_{I_3^{ X} I_3} \left ( \frac{M_0}{3} \right )  \langle n_\rho \vert \rho^2 \vert n_\rho \rangle \,  \cr
 &\times& \vec{q}^2 \left ( \delta^{ia} - \frac{q^i q^a}{\vec{q}^2} \right )  \chi^\dagger_{s_{ X}}(\vec{p_{ X}}) \sigma_a  \chi_s (\vec{p})  \alpha F^2_{B B_{ X}}(\vec{q}^2) \, , \label{SSCurrentBreitNegParity}
\eeqa
when the final state has negative parity. In (\ref{SSCurrentBreitPosParity}) and (\ref{SSCurrentBreitNegParity})
we used the definitions
\beqa
\xi &=& \left (\frac{1}{2E} \right ) \left ( \frac{f}{f_X} \right ) \left [ f_X^2 + \frac{\vec{p}^2}{f^2} (2x-1) \right ] \, , \cr
 \left ( \frac{M_0}{3} \right ) \langle n_\rho \vert \rho^2 \vert n_\rho \rangle &=&
\frac{1}{\sqrt{6} M_{KK} } \left [ 1 + 2 \sqrt{ 1 + \frac{N_c^2}{5} }  \right ] =: \frac{g_{I=1}}{4 m_B}  \, , \label{xidef}
\eeqa
and  $\alpha$ was defined in (\ref{defalpha}).

\subsection{Dirac and Pauli form factors in the Sakai-Sugimoto model}

We are going to use the holographic prescription
\beqa
 \eta_\mu \langle p_{  X} , B_{  X} , s_{  X}  \vert J_V^{\mu,a} (0) \vert p , B, s \rangle
= \eta_\mu \langle p_{  X} , B_{  X} , s_{  X}  \vert J_{V (SS)}^{\mu,a} (0) \vert p , B, s \rangle \, ,  \label{holprescription}
\eeqa
where $\eta_\mu = (\eta_0 , \vec{\eta})$ is the polarization of the photon and we choose to work with transverse photons satisfying the relation $\eta_\mu q^\mu=0$
in order to avoid the discussion of current anomalies.

Using (\ref{holprescription}) we can compare, the kinematic currents (\ref{finalcurrent0}), (\ref{finalcurrenti}),
(\ref{chfinalcurrent0}) and (\ref{chfinalcurrenti}) with the Sakai-Sugimoto currents
(\ref{SSCurrentBreitPosParity}) and (\ref{SSCurrentBreitNegParity}).
For positive parity resonances we get
\beqa
F^{D,0}_{B B_{  X}}(q^2) &=& \left [ \frac{\xi \alpha + \beta \alpha \frac{q^2 }{4 M_0}  }{\alpha^2 + \beta^2 q^2} \right ] N_c F^1_{B B_{ X}}(q^2) \, , \cr
F^{P,0}_{B B_{  X}}(q^2) &=& - \frac{1}{\kappa_B} \left [ \frac{\beta \xi - \frac{\alpha^2}{4 M_0}  }{\alpha^2 + \beta^2 q^2 } \right ]  N_c F^1_{B B_{ X}}(q^2) \, , \cr
F^{D,3}_{B B_{  X}}(q^2) &=& \left [ \frac{\xi \alpha + \beta \alpha q^2  \left ( \frac{M_0}{3} \right ) \langle \rho^2 \rangle  }{\alpha^2 + \beta^2 q^2} \right ]   F^1_{B B_{ X}}(q^2) \, , \cr
F^{P,3}_{B B_{  X}}(q^2) &=& - \frac{1}{\kappa_B} \left [ \frac{\beta \xi - \alpha^2 \left ( \frac{M_0}{3} \right ) \langle \rho^2 \rangle }{\alpha^2 + \beta^2 q^2 } \right ]   F^1_{B B_{ X}}(q^2) \, ,
\eeqa
where $\alpha$ and $\beta$ are given in (\ref{defalpha}), (\ref{defbeta}) and $\xi$ is given in (\ref{xidef}).
For negative parity resonances, we can write
\beqa
\tilde F^{D,0}_{B B_{  X}}(q^2) &=& x \left ( \frac{q^2}{2M_0} \right ) \left [ \frac{ \hat \beta \alpha}{\hat \alpha^2 + \hat \beta^2 q^2} \right ] N_c F^2_{B B_{ X}}(q^2) \, , \cr
\tilde F^{P,0}_{B B_{  X}}(q^2) &=& x \left ( \frac{1}{2 M_0 \kappa_B} \right ) \left [ \frac{ \hat \alpha \alpha}{\hat \alpha^2 + \hat \beta^2 q^2} \right ] N_c F^2_{B B_{ X}}(q^2) \, , \cr
\tilde F^{D,3}_{B B_{  X}}(q^2) &=& 2 x \left [ \frac{ \frac{M_0}{3} \langle \rho^2 \rangle  \hat \beta \alpha q^2  - \hat \alpha \xi }{\hat \alpha^2 + \hat \beta^2 q^2} \right ] F^2_{B B_{ X}}(q^2) \, , \cr
\tilde F^{P,3}_{B B_{  X}}(q^2) &=&  2 x \left ( \frac{1}{\kappa_B} \right )\left [ \frac{ \frac{M_0}{3} \langle \rho^2 \rangle  \hat \alpha \alpha  + \hat \beta \xi }{\hat \alpha^2 + \hat \beta^2 q^2} \right ]
 F^2_{B B_{ X}}(q^2) \, ,
\eeqa
where $\hat \alpha$ and $\hat \beta$ are given in (\ref{hatalpha}) and (\ref{hatbeta}), respectively.
We relegate the details of the large $\lambda$ expansions relevant to get the dominant contribution to the
form factors in the non-elastic case to appendix B. The large $\lambda$ limit in the elastic case, corresponding to
$n_{\scst  X}=0$, $m_{B_{\scst  X}}=m_B$ was already considered in \cite{Hashimoto:2008zw}.
In the non-elastic case, the positive parity resonances correspond to $n_{B_X}={2,4,6,\dots}$.
In this case, we get in the large $\lambda$ limit \cite{Bayona:2011xj} :
\beqa
F^{D,0}_{B B_{  X}}(q^2) &=& \left [ \frac{m_B}{E} + {\cal O} \left ( \frac{1}{ \lambda N_c} \right )  \right ]
  N_c F^1_{B B_{ X}}(q^2) \, , \cr
F^{P,0}_{B B_{  X}}(q^2) &=&  \left [  \frac{g_{I=0}}{2} -\frac{m_B}{E} + {\cal O}
 \left ( \frac{1}{ \lambda N_c} \right ) \right ]  N_c F^1_{B B_{ X}}(q^2) \, , \cr
F^{D,3}_{B B_{  X}}(q^2) &=& \left [ \frac{m_B}{E} + {\cal O}
\left ( \frac{1}{\lambda} \right ) \right ]   F^1_{B B_{ X}}(q^2) \, , \cr
F^{P,3}_{B B_{  X}}(q^2) &=& \frac{g_{I=1}}{2} \left[ 1 + {\cal O} \left ( \frac{1}{ \lambda N_c} \right )
\right  ]   F^1_{B B_{ X}}(q^2) \, .
\eeqa
For negative parity resonances, we have $n_{B_X}={1,3,5,\dots}$. Using the expansions in appendix B is not
difficult to show that in the large $\lambda$ limit the form factors reduce to
\beqa
\tilde F^{D,0}_{B B_{  X}}(q^2) &=& \frac{q^2}{4E} g_{I=0} \left [ 1 + {\cal O} \left ( \frac{1}{ \lambda N_c} \right ) \right ]  N_c F^2_{B B_{ X}}(q^2) \, , \\
\tilde F^{P,0}_{B B_{  X}}(q^2) &=& \left ( \frac{1}{x} \right ) \frac{q^2}{4 E} g_{I=0} \left [ 1 + {\cal O} \left ( \frac{1}{ \lambda N_c} \right ) \right ]  N_c F^2_{B B_{ X}}(q^2) \, , \\
\tilde F^{D,3}_{B B_{  X}}(q^2) &=& \frac{q^2}{4E} g_{I=1} \left [ 1 + {\cal O} \left ( \frac{1}{ N_c} \right ) \right ] F^2_{B B_{ X}}(q^2) \, , \\
\tilde F^{P,3}_{B B_{  X}}(q^2) &=& \left ( \frac{1}{x} \right ) \frac{q^2}{4E} g_{I=1} \left [ 1 + {\cal O} \left ( \frac{1}{ N_c} \right ) \right ]  F^2_{B B_{ X}}(q^2) \, .
\label{FDFPhologneg}
\eeqa

\section{Numerical results for negative parity baryons}

We present in this section our numerical results for the negative parity baryons. These
include the wave functions, Dirac and Pauli form factors, helicity amplitudes and their
contribution to the proton structure function. We are using the Sakai-Sugimoto parameters
$M_{KK}=949$MeV and $\kappa = 7.45 \times 10^{-3}$ \cite{Sakai:2005yt}.
We also choose  $\tilde M_0 = 940 \text{MeV}$, for phenomenological reasons.

\subsection{Baryon wave functions}

\begin{figure}[ht]
\begin{center}
\includegraphics[width=.38 \textwidth]{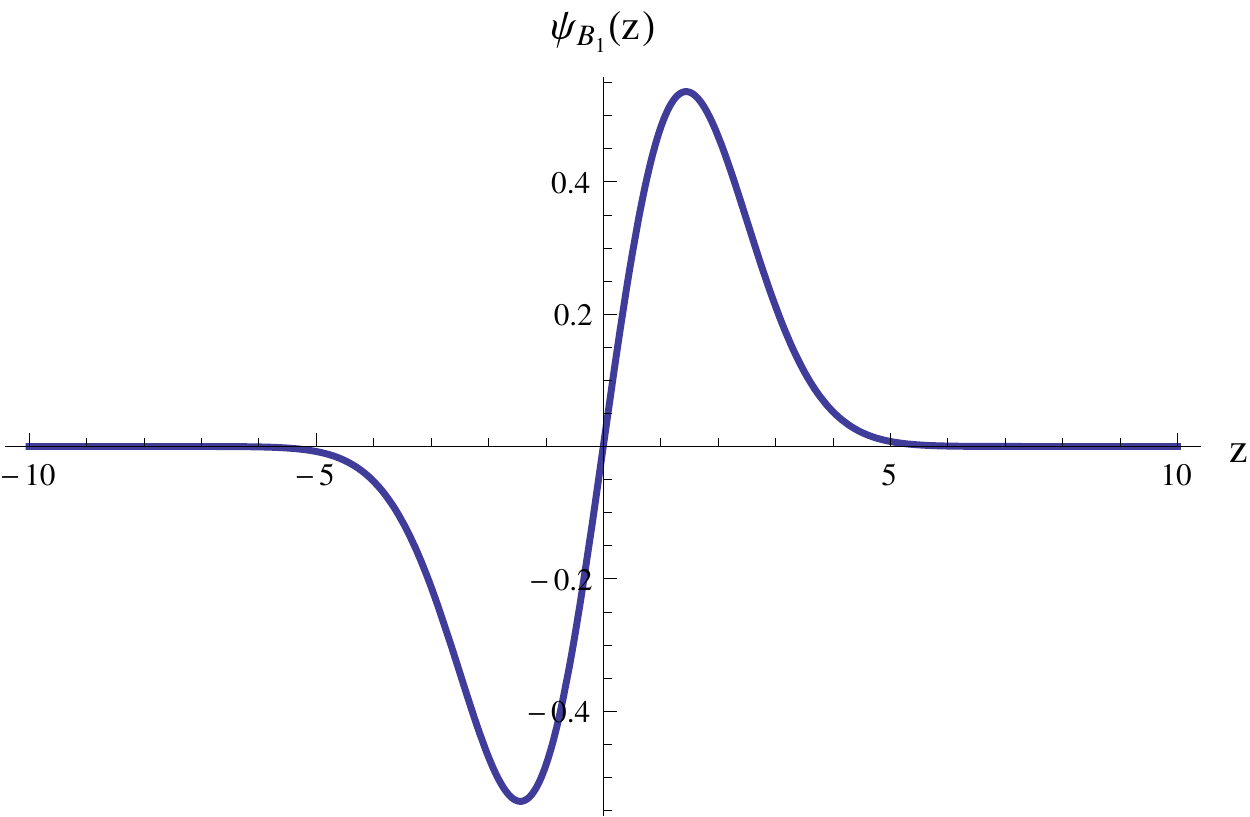}
\includegraphics[width=.38 \textwidth]{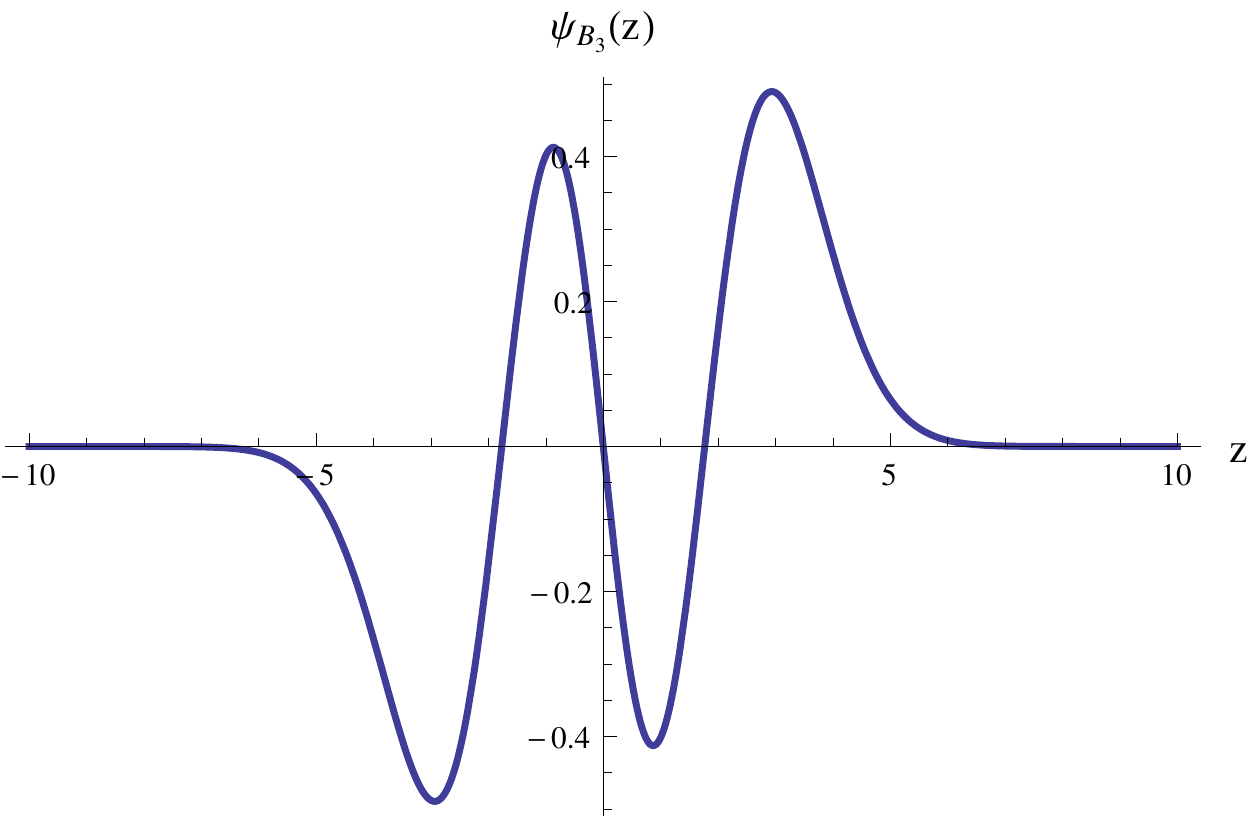}
\includegraphics[width=.38 \textwidth]{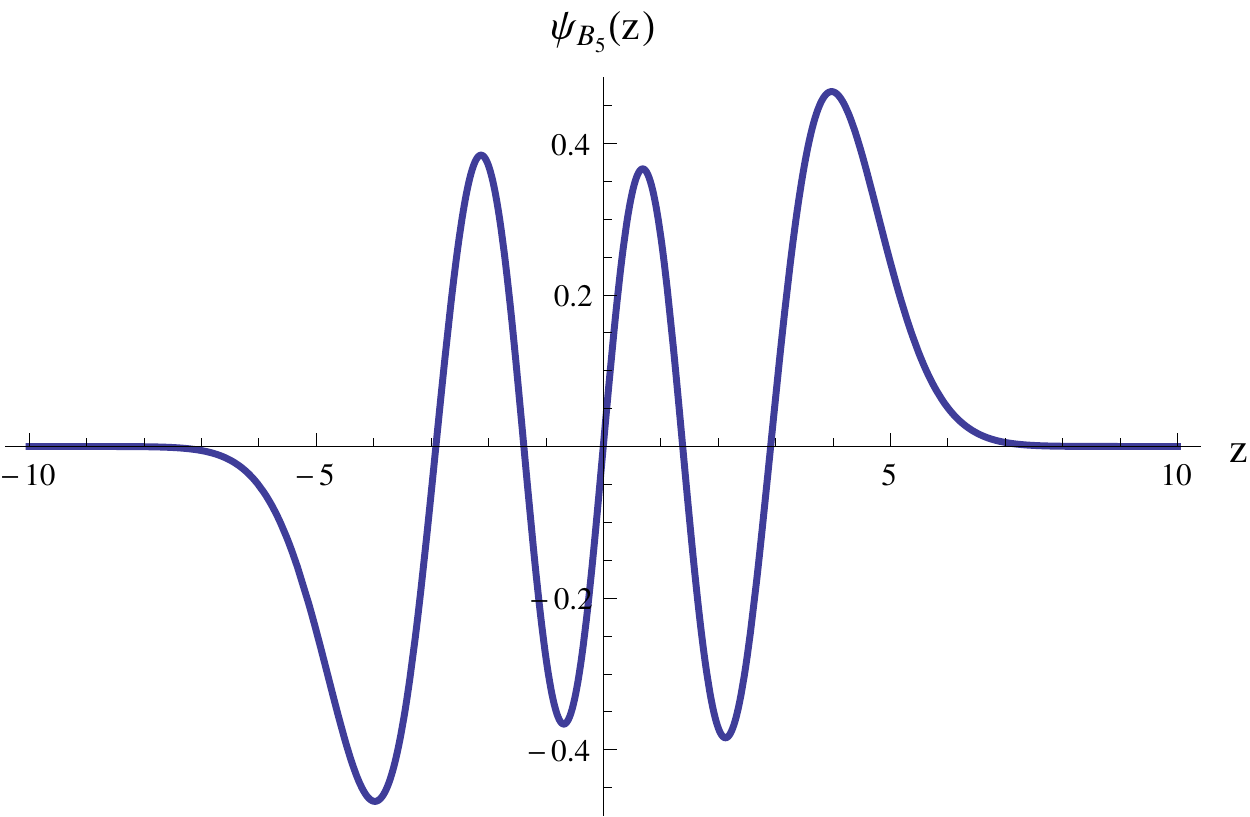}
\includegraphics[width=.38 \textwidth]{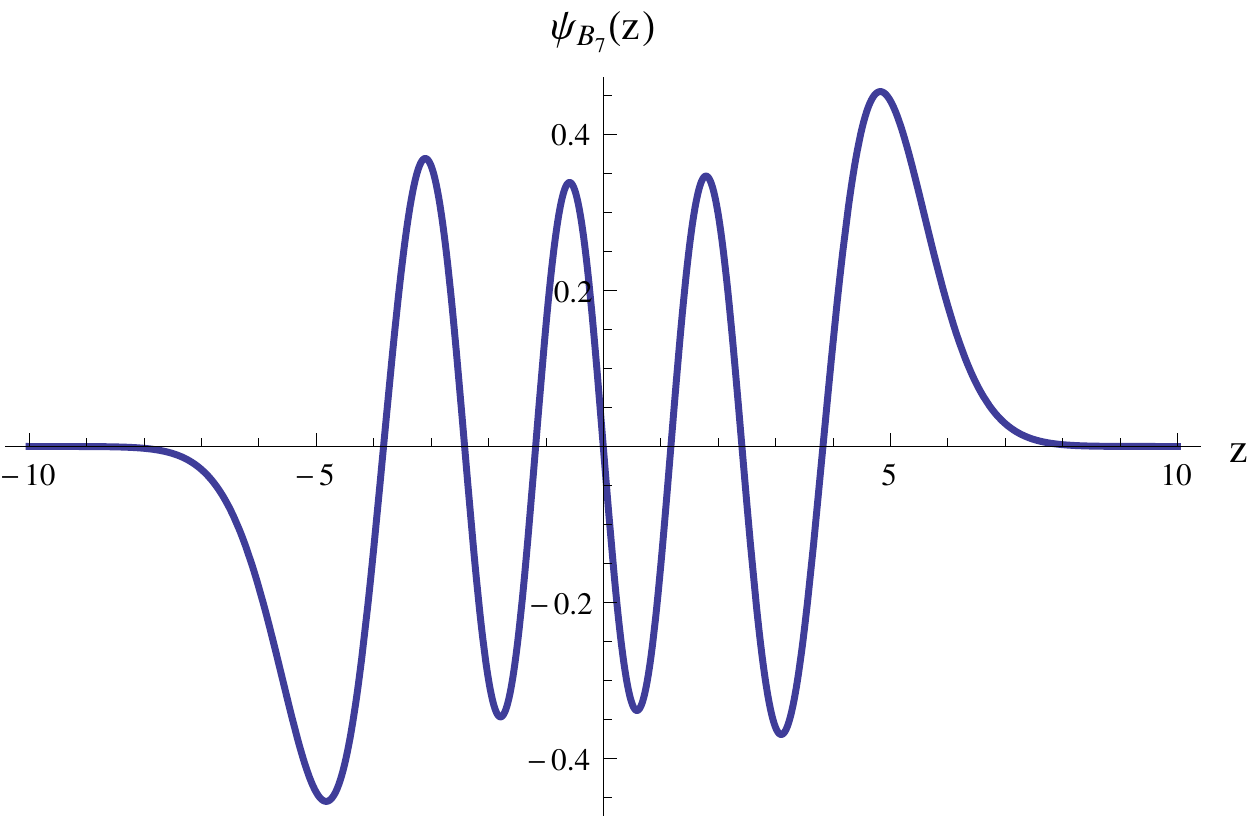}
\includegraphics[width=.38 \textwidth]{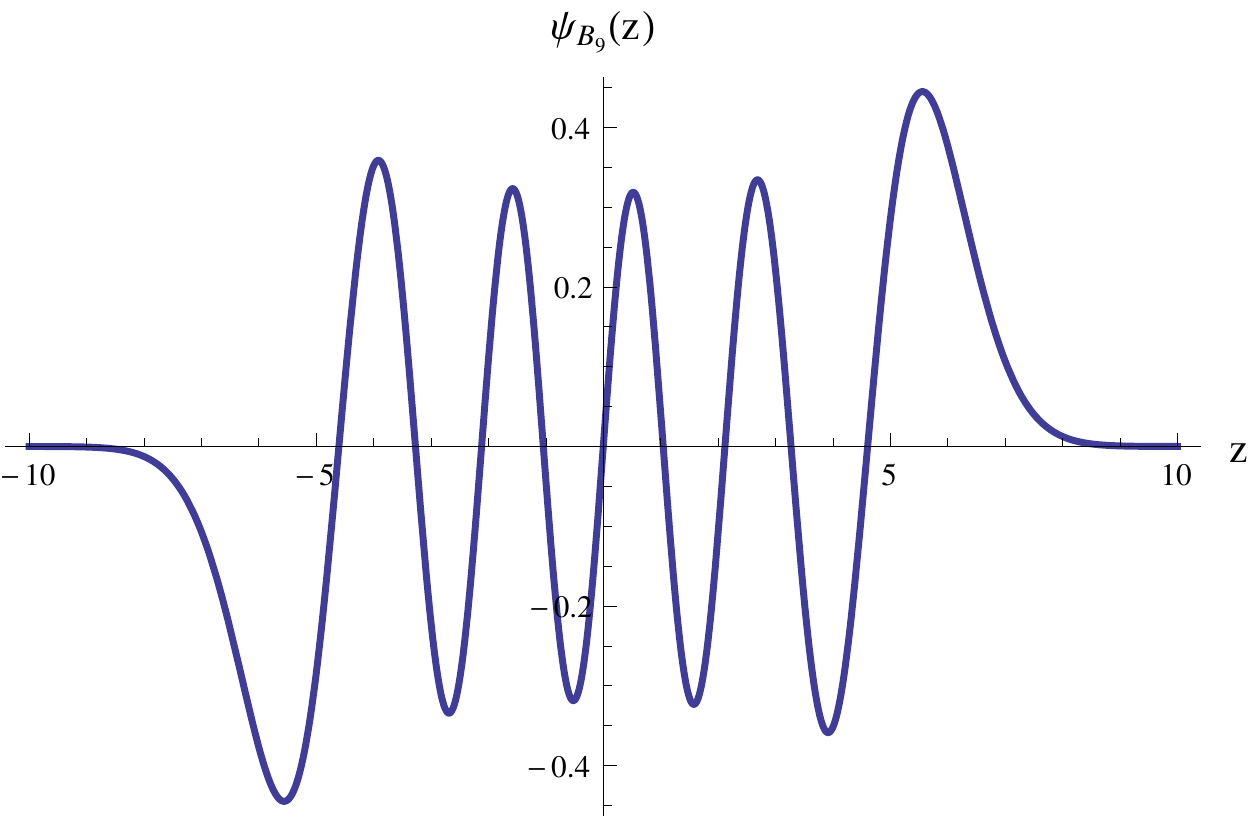}
\includegraphics[width=.38 \textwidth]{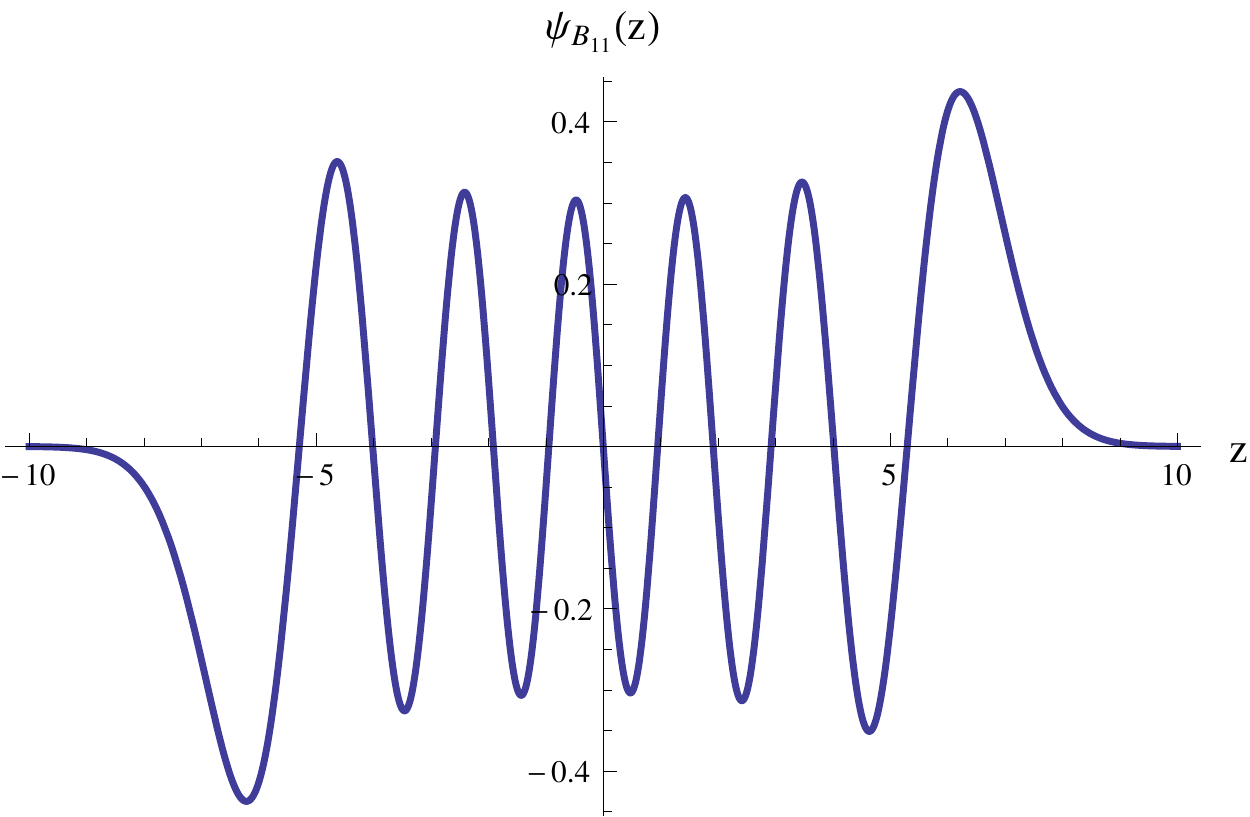}
\end{center}
\caption{(Normalized) wave functions $ \Psi_{B_{2k - 1}}(z)$ for the first six parity odd baryon states.}
\label{fig:baryonwavefct}
\end{figure}

First we present in fig. \ref{fig:baryonwavefct} the results for the wave functions of the first excited
baryons with negative parity. These wave functions have quantum numbers $B_n = (1 , +1/2, 0, n)$ with
$n= 2 k - 1$ and are odd functions in the radial coordinate $z$.
Table \ref{tab:baryonmasses} shows the mass spectrum of the first negative parity baryonic resonances. The spectrum
of positive parity baryonic resonances can be found in \cite{Bayona:2011xj}.
\begin{table}[ht]
\begin{center}
 \begin{tabular}{|c||c|c|c|c|c|c|c|c||}
 \hline
 $n$ &  1 & 3 & 5 & 7 & 9 & 11 & 13 & 15  \\ \hline
 $m_{B_n}$/GeV & 1.715 & 3.265 & 4.814 & 6.364 & 7.914 & 9.463  & 11.013 & 12.563  \\\hline
\end{tabular}
\end{center}
\caption{Some numerical values for the masses of negative parity baryon states}\label{tab:baryonmasses}
\end{table}

\subsection{Dirac and Pauli Form factors}
\label{subsec:formfactors}
In the previous section we extracted from holography the Dirac and Pauli form factors that describe the production of
negative parity baryons. Interestingly, our results (\ref{FDFPhologneg}) show that the Dirac and Pauli form factors
depend on only one form factor $F^2_{B B_{ X}}(\vec{q}^2)$ defined by (\ref{masterformfactors}). This is a feature
that has also appeared in previous holographic approaches to electromagnetic scattering\footnote{See
\cite{BallonBayona:2009ar,Bayona:2010bg} for a similar result for vector meson form factors in the holographic approach.}. The form factor $F^2_{B B_{ X}}(\vec{q}^2)$
in (\ref{masterformfactors}) can be written as
\beqa
F^2_{B B_{ X}}(\vec{q}^2) = \sum_n \frac{g_{v^n} g_{v^n B_0 B_X} }{\vec{q}^2 + \lambda_{2n-1}}
\eeqa
where
\beqa
g_{v^n B_0 B_X} := \frac{1}{\lambda_{2n-1}}
\langle n_{B_{ X}} \vert \partial_Z \psi_{2n-1}(Z)  \vert n_B \rangle \,,
\eeqa
are the effective couplings between a vector meson, a negative parity baryon and the proton. We show in
table \ref{tab:barCC} our numerical results for these effective couplings. Identifying the first
negative parity resonance with the experimentally observed $S_{11}(1535)$, our numerical result
for the coupling constant $g_{v1 B_0 B_1} = -1.889$ should be useful to
describe the decay of $S_{11}(1535)$ into a $\rho$ meson and a proton. This result is compatible with
recent analysis from experimental data \cite{Xie:2008ts} where $0.79 < |g_{v1 B_0 B_1}| < 2.63 $. The vector meson
squared masses $\lambda_{2n-1}$ and decay contants $g_{v^n}$ are also shown in table \ref{tab:barCC} .
\begin{table}[ht]
\begin{center}
\begin{tabular}{|c||c|c|c|c|c|c|c|c||}
\hline
$n$ & 1 & 2 & 3 & 4 & 5 & 6 & 7 & 8  \\\hline\hline
$\lambda_{2n-1}$ & 0.6693 & 2.874 & 6.591 & 11.80 & 18.49 & 26.67 & 36.34 & 47.49  \\\hline
$\frac{g_{v^n}}{\sqrt{\kappa}M_{KK}^2}$ & 2.109 & 9.108 & 20.80 & 37.15 & 58.17 & 83.83 & 114.2 & 149.1 \\\hline
$g_{v^nB_0B_1}$  & -1.889 &  1.182 & -0.562 &  0.1381 & 0.04057 & -0.05213 & 0.01239 & 0.009893  \\\hline
$g_{v^nB_0B_3}$  & 1.038  & -0.841 & 0.6132 & -0.3325 & 0.08703 & 0.04209  & -0.05382 & 0.01409    \\\hline
$g_{v^nB_0B_5}$  & -0.6432 & 0.5892 & -0.5217 & 0.3802 & -0.1907 & 0.02706 & 0.05097 & -0.04458   \\\hline
$g_{v^nB_0B_7}$  & 0.429  & -0.4239 & 0.4223 & -0.3644 & 0.2416 & -0.09421 & -0.01629 & 0.05276  \\\hline
$g_{v^nB_0B_9}$  & -0.3005 & 0.3132 & -0.3386 & 0.3273 &  -0.2571 & 0.1417 & -0.0266 & -0.0404    \\\hline\hline
\end{tabular}
\end{center}
\caption{Coupling constants between vector mesons and baryons when the initial state is the proton and the final state has negative parity.}\label{tab:barCC}
\end{table}

\begin{figure}[h!]
\begin{center}
\includegraphics[width=.38 \textwidth]{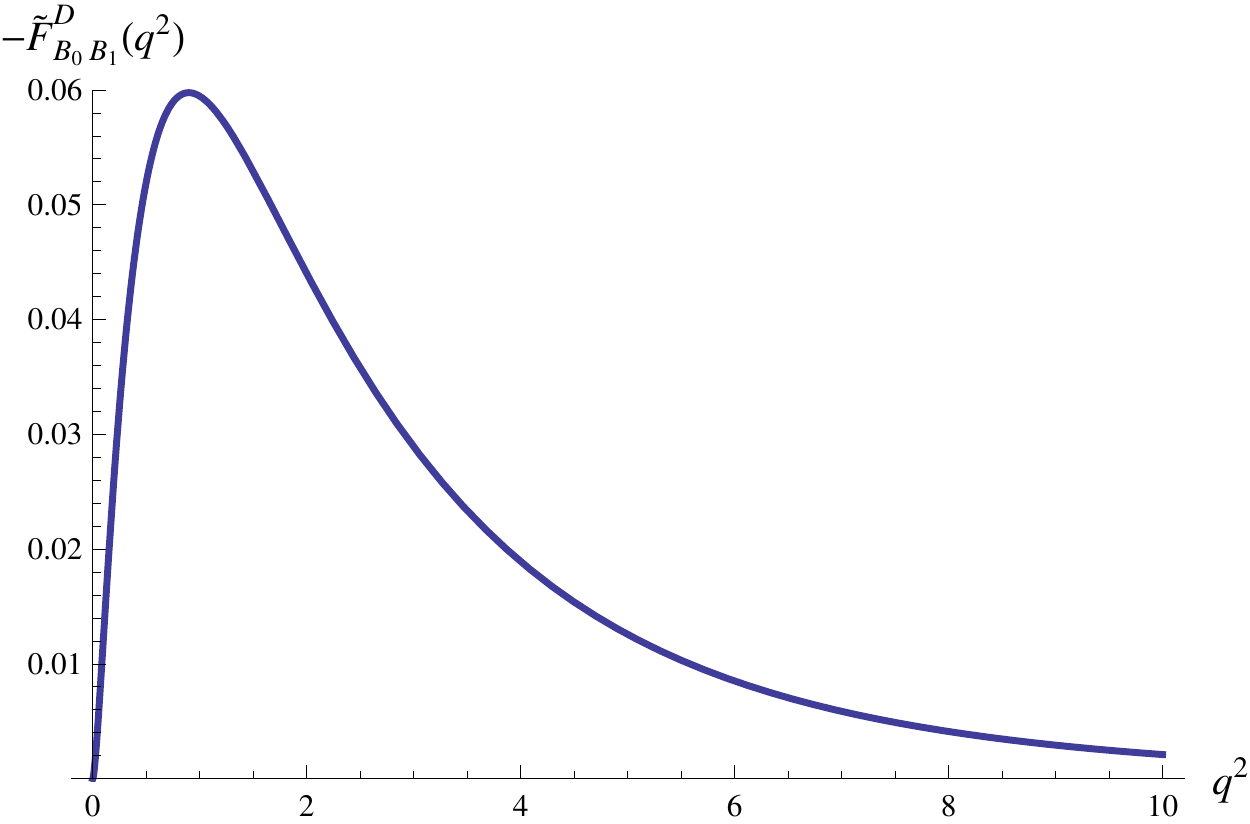}
\includegraphics[width=.38 \textwidth]{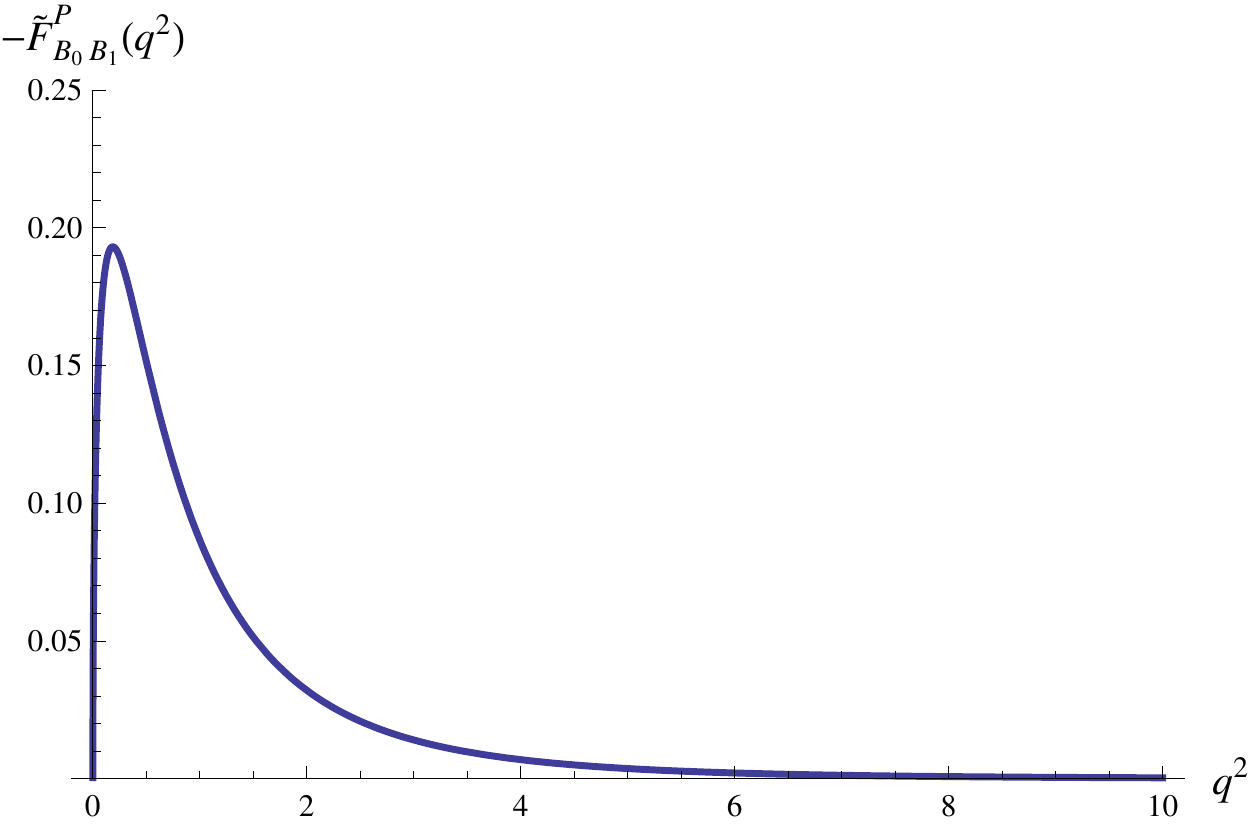}
\includegraphics[width=.38 \textwidth]{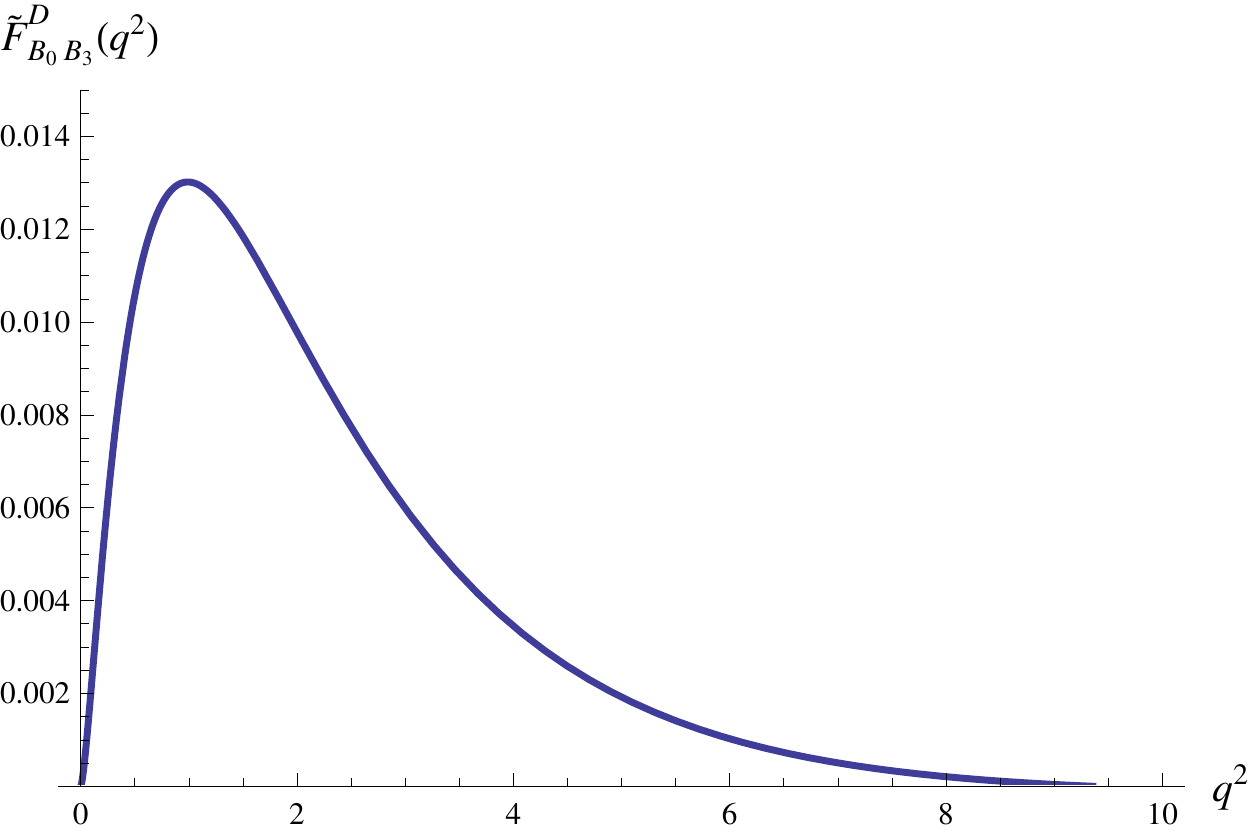}
\includegraphics[width=.38 \textwidth]{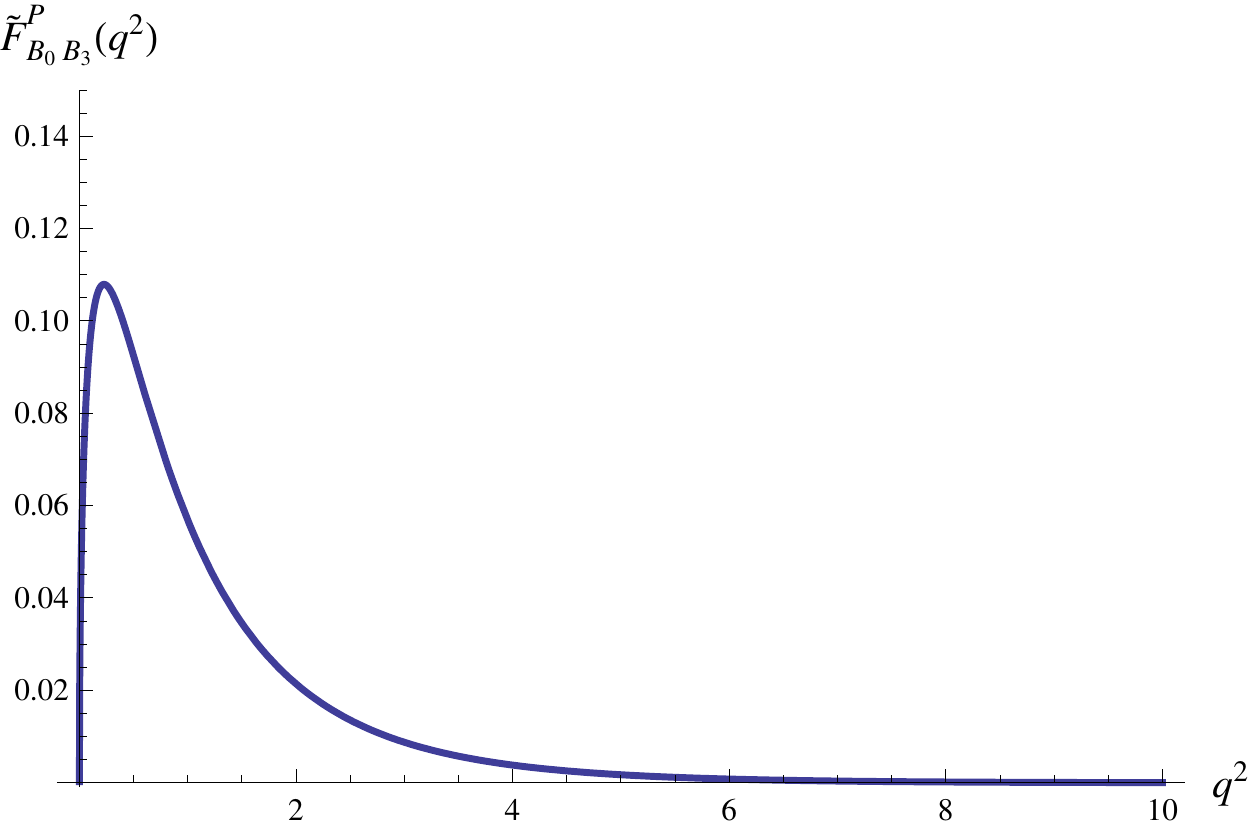}
\includegraphics[width=.38 \textwidth]{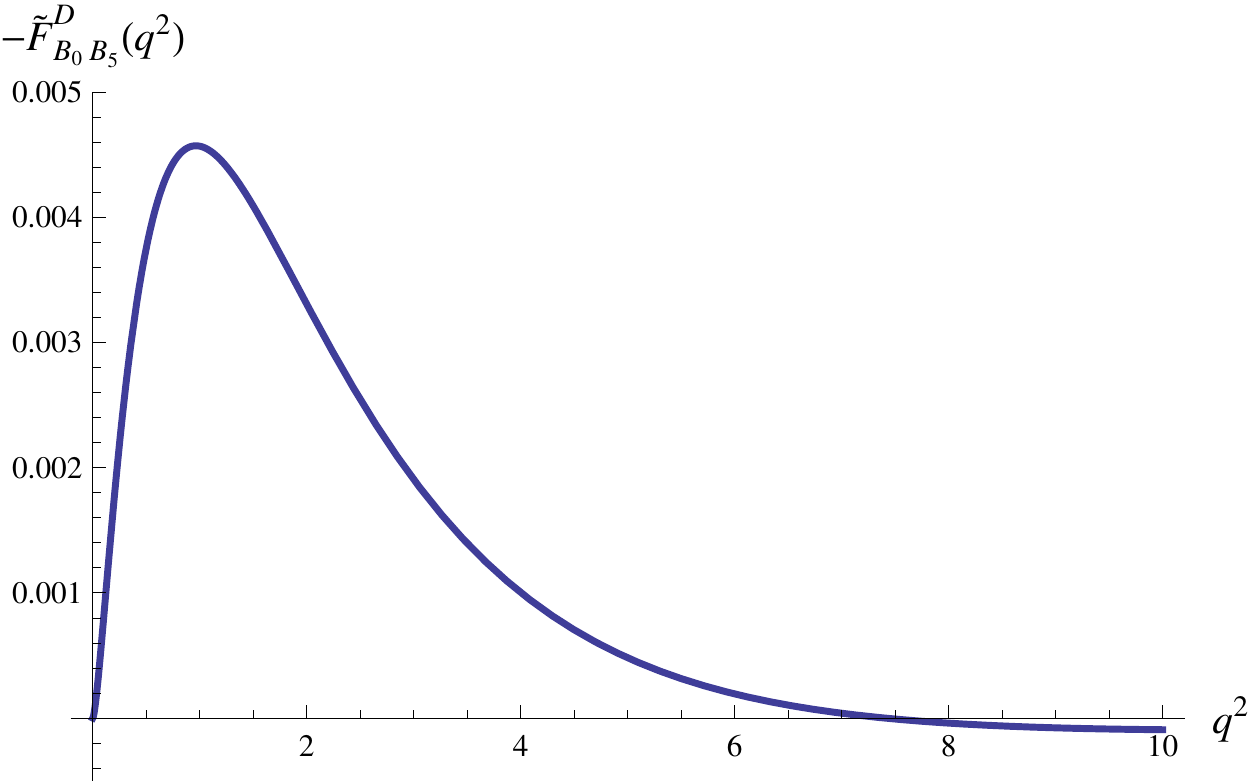}
\includegraphics[width=.38 \textwidth]{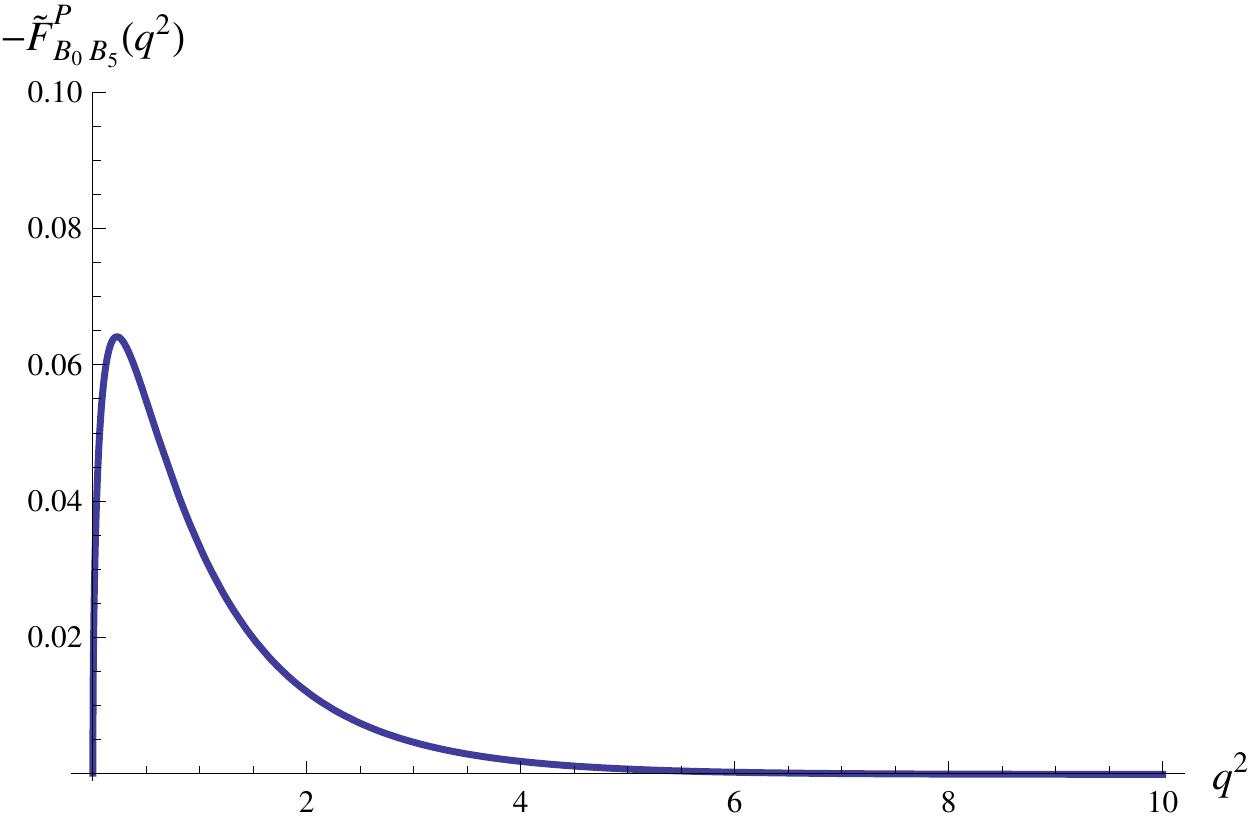}
\end{center}
\caption{Dirac and Pauli form factors $\tilde F^{D,P}_{B_0 B_{2j - 1}}(q^2)$ for the first three negative parity
baryon states. The momentum transfer $q^2$ is given in $(\text{GeV})^2$.}
\label{fig:FF}
\end{figure}

The Dirac and Pauli form factors depend on the magnetic $g_I$ factors whose numerical values in the Sakai-Sugimoto model are
given by
\beqa
g_{I=0} \approx 1.684 \, , \qquad g_{I=1} \approx 7.031 \,. \label{gIfactors}
\eeqa
Using (\ref{gIfactors}) and our results for the couplings, masses and decay constants (shown in table \ref{tab:barCC})
we can calculate the Dirac and Pauli form factors describing the production of negative parity baryon states.
We show our results for the first three excited states in figure \ref{fig:FF}. As a general feature,
the form factors go to zero as $q^2 \to 0$, reach a maximum and then decay for large $q^2$.
Note that some of the form factors are non-positive.

\subsection{Helicity amplitudes: comparison with JLab-CLAS data}

In the large $\lambda$ limit, the transverse helicity amplitudes take the form
\beqa
\tilde G^+_{B B_{\scst  X}}(q^2) &\approx& -\sqrt{2} \left [ \tilde F^D_{B B_{\scst  X}}(q^2)
+ \frac{m_{B_X} - m_B}{2 m_B}  \tilde F^P_{B B_{\scst  X}}(q^2) \right ] \, , \cr
\tilde {\cal A}^{1/2}_{B B_{\scst  X}}(q^2) &\approx& \frac{e}{\sqrt{2 (m_{B_X} - m_B)}} \tilde G^+_{B B_{\scst  X}}(q^2) \, ,
\eeqa

Unfortunately, in the large $\lambda$ limit we cannot say too much about the longitudinal
helicity amplitudes because we obtain
\beqa
\tilde G^0_{B B_{\scst  X}}(q^2) &\approx&  \sqrt{q^2} \left [ \frac{m_{B_X} - m_B}{q^2} \tilde F^D_{B B_{\scst  X}}(q^2)
- \frac{1}{2 m_B} \tilde F^P_{B B_{\scst  X}}(q^2) \right ]  \approx 0 \,  , \cr
\tilde {\cal S}^{1/2}_{B B_{\scst  X}}(q^2) &\approx& e \sqrt{ \frac{m_B}{q^2}} \tilde G^0_{B B_{\scst  X}}(q^2) \approx 0  \,.
\eeqa

This result seems to be consistent with the fact that the experimental data available for these helicity amplitudes
indicates a strong contribution from meson clouds \cite{Aznauryan:2009mx}. This kind of effect necessitates the investigation of
loop corrections of order $1/ \lambda$ in electromagnetic scattering. The $1/\lambda$ corrections would not only modify our results but also
the standard results on the elastic electromagnetic form factors\footnote{See \cite{Cherman:2009gb} for a discussion
regarding pion loop corrections in baryon electromagnetic form factors.}.  

\begin{figure}[h!]
\begin{center}
\includegraphics[width=.48 \textwidth]{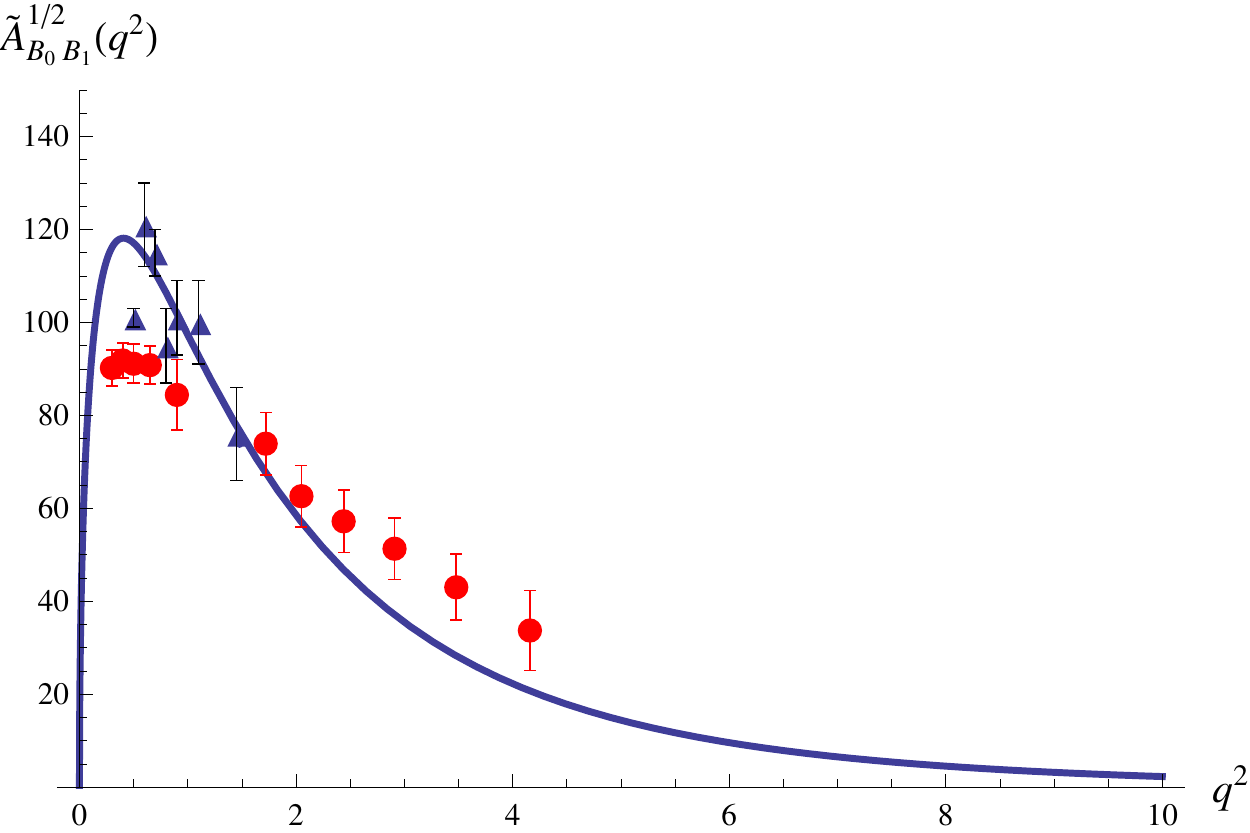}
\end{center}
\caption{Helicity amplitude $\tilde {\cal A}^{1/2}_{B B_{\scst  1}}(q^2)$
(in units $10^{-3}(\text{GeV})^{-1/2}$) plotted versus $q^2$ in $(\text{GeV})^2$.
The JLAB-CLAS experimental data (red dots) was taken from ref. \cite{Aznauryan:2009mx}, while the EBAC bare amplitude results (blue triangles)  were taken from \cite{JuliaDiaz:2009ww, Ramalho:2011ae}. }
\label{fig:helamp}
\end{figure}

Some of the meson cloud contributions to the helicity amplitudes $\tilde {\cal A}^{1/2}$ and $\tilde {\cal S}^{1/2}$ for the resonance $S_{11}(1535)$ were calculated by 
the EBAC group \cite{JuliaDiaz:2009ww}, fitting the dynamical coupled-channel model \cite{Matsuyama:2006rp} with experimental data. The EBAC result \cite{JuliaDiaz:2009ww}  was displayed nicely in \cite{Ramalho:2011ae}, where the authors removed the meson cloud contributions from the dressed helicity amplitudes and presented the EBAC bare helicity amplitudes.  In figure \ref{fig:helamp}, we present our result for the transverse helicity amplitude $\tilde {\cal A}^{1/2}_{B B_{1}}(q^2)$ for the first negative parity resonance and also show the EBAC results \cite{JuliaDiaz:2009ww, Ramalho:2011ae} and recent experimental data from the JLAB-CLAS collaboration \cite{Aznauryan:2009mx} for comparison. As discussed in section \ref{subsec:formfactors}, this resonance can be identified with the experimentally observed $S_{11}(1535)$.  
Despite the limitations of our model (the large $\lambda$ limit), we find good agreement with the EBAC results for bare helicity amplitudes and reasonable agreement with JLAB-CLAS experimental data. This is to be expected since the EBAC data is available for $q^2\leq 1.5(\text{GeV})^2$, which coincides with the regime of validity of the Sakai-Sugimoto model where we expect our results to be reliable, whereas the JLAB-CLAS data extends to higher $q^2$ beyond the regime of validity. Furthermore, the EBAC and JLAB-CLAS results clearly demonstrate the importance of $1/\lambda$-corrections from meson cloud contributions in the non-perturbative regime, which means that we need to go beyond tree level to get a better than qualitative agreement with experimental results for helicity amplitudes. Nevertheless, this is a very encouraging result in view of our long-term project of investigating resonance production in holographic models.

\begin{figure}[h!]
\begin{center}
\includegraphics[width=.48 \textwidth]{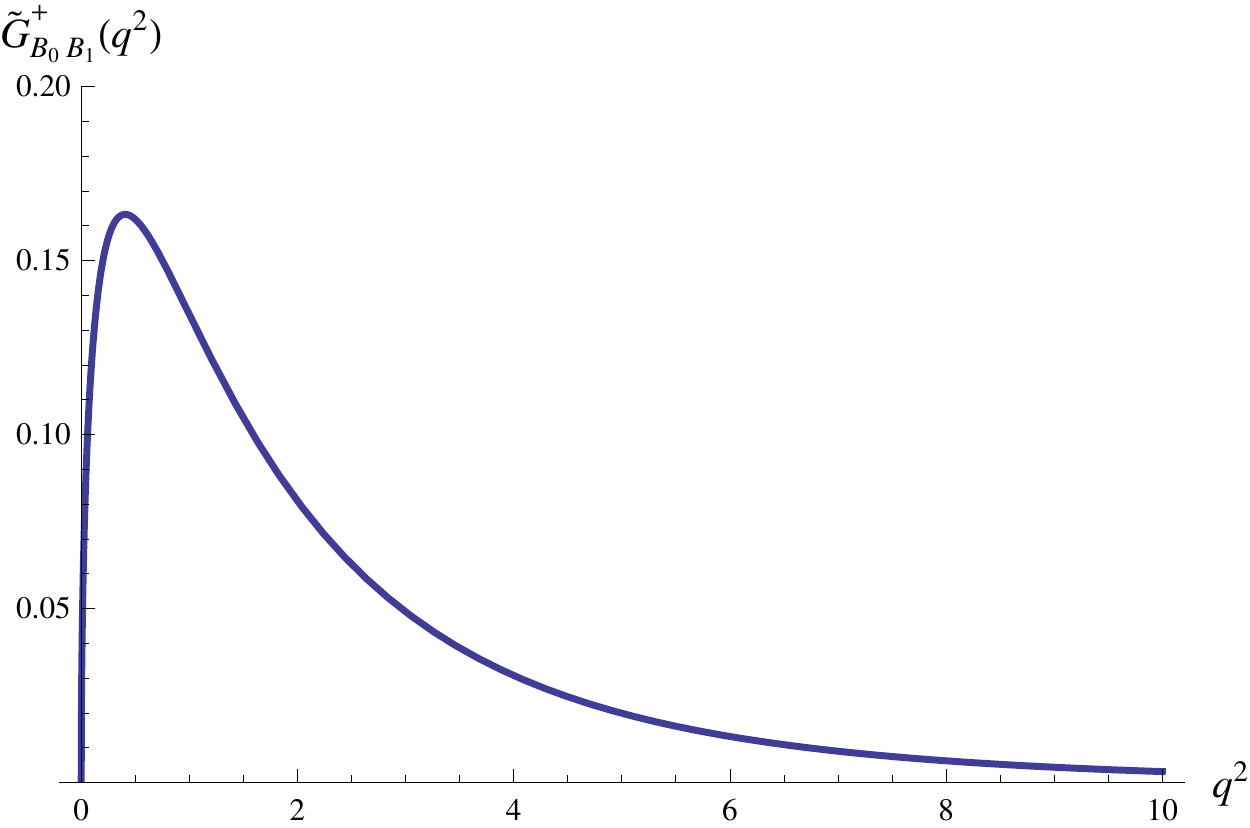}
\end{center}
\caption{Helicity amplitude  $\tilde G^{+}_{B_0 B_1}(q^2)$ 
(in units $10^{-3}(\text{GeV})^{-1/2}$) plotted versus $q^2$ in $(\text{GeV})^2$.}
\label{fig:Ghelamp}
\end{figure}

In figure \ref{fig:Ghelamp} we show our results for $\tilde G^+_{B B_{1}}(q^2)$.

\noindent
Note that, in our model, $\tilde {\cal A}^{1/2}_{B B_{\scst X}}(q^2) \rightarrow 0$ as $q^2 \rightarrow 0$, contrary to the experimental results reported in \cite{Beringer:1900zz}, p. 1155,
\begin{equation}
\tilde {\cal A}^{1/2 \,\text{exp.}}_{B B_{\scst 1}}(q^2) = 0.09 \pm 0.03 \,(\text{GeV})^{-1/2}. 
\end{equation}
This can be understood from the fact that the baryons in our model are very massive and thus stable in the large $\lambda$ limit. 
Therefore the currents that we construct do not account for the decay of baryonic resonances. 

\subsection{The proton structure function}

\subsubsection{A first approximation}

Assuming approximate continuity of the mass distribution, we can now approximate the delta distributions in the following way:
\begin{eqnarray}
\sum_{B_{X}} \delta [ m_{B_{X}}^2 -s ]
&=& \sum_{n} \delta [ m_{n}^2 -m_{\bar n}^2 ]
= \int dn
\left[\left| \frac{\partial m_n^2}{\partial n} \right|\right]^{-1}
\delta(n-\bar n)\cr
&=&\left[\left| \frac{\partial m_n^2}{\partial n} \right|\right]_{n=\bar n}^{-1} =: f(\bar n), \label{approxDelta}
\end{eqnarray}
with the definition
\beqa\label{eq:s}
s := -(p+q)^2 = m_{B_0}^2 + q^2 ( \frac 1x -1).
\eeqa
Therefore we have to evaluate the Regge trajectory of the baryon spectrum in order to calculate $\frac{\partial m_n^2}{\partial n}$.
We find from (\ref{baryonmass})
\beqa
\frac{\partial m_n^2}{\partial n} = \left(\frac{4}{\sqrt{6}}{\wt M}_0 M_{KK}+ \frac{4}{3} n M_{KK}^2\right),
\eeqa
where ${\wt M}_0$ can be chosen to match, e.g. the proton mass $m_{B_0}$ and $n:= n_z$.\\
Using the approximation (\ref{approxDelta}) we get in the large $\lambda$ limit the structure functions
\beqa
\tilde F_1 (q^2 ,x)
&\approx & f(\bar n) m_B^2 ( \tilde G^+_{B B_{\bar n}}(q^2) )^2 \, , \cr
\tilde F_2 (q^2 ,x) &\approx&   f(\bar n)  \left (\frac{q^2}{2x} \right ) \left (1 + \frac{q^2}{4 m_B^2 x^2} \right )^{-1}
(\tilde G^+_{B B_{\bar n}}(q^2))^2 \,. \label{naiveF12}
\eeqa

\begin{figure}[h]
\begin{center}
\includegraphics[width=.38 \textwidth]{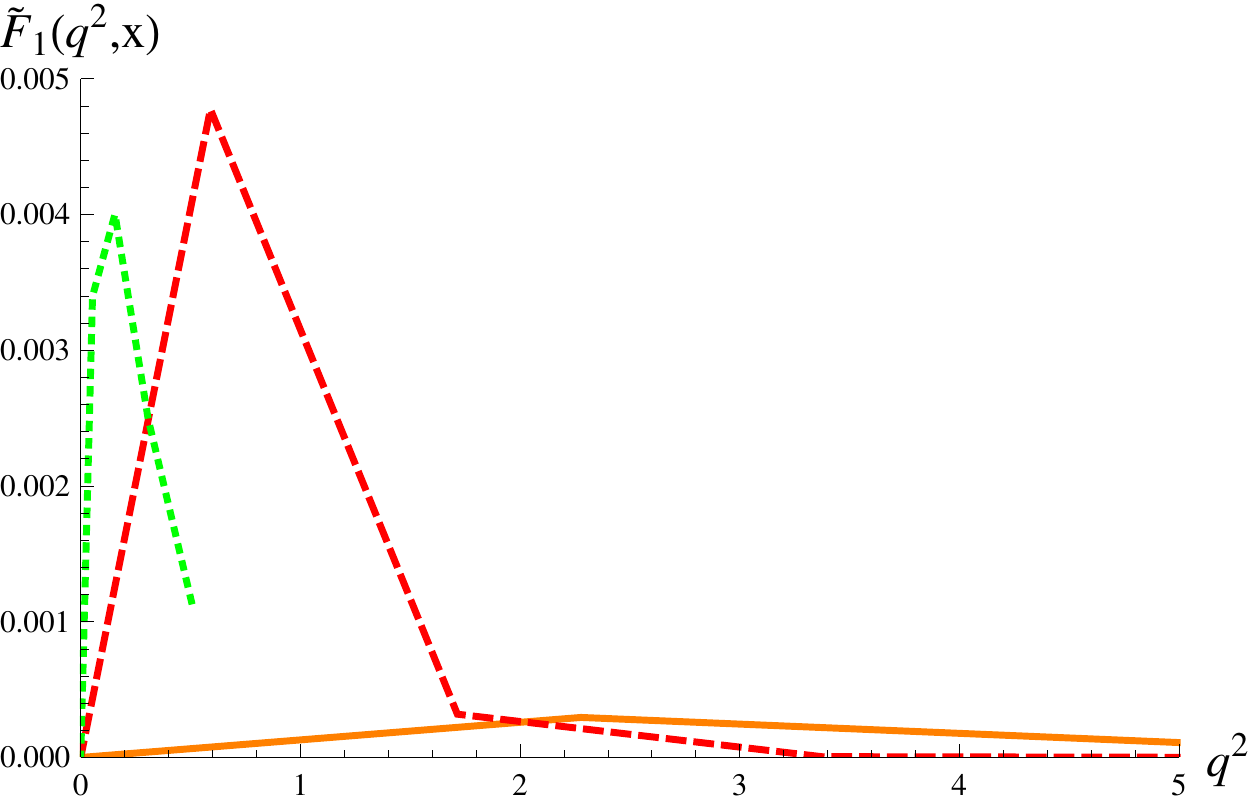}
\includegraphics[width=.38 \textwidth]{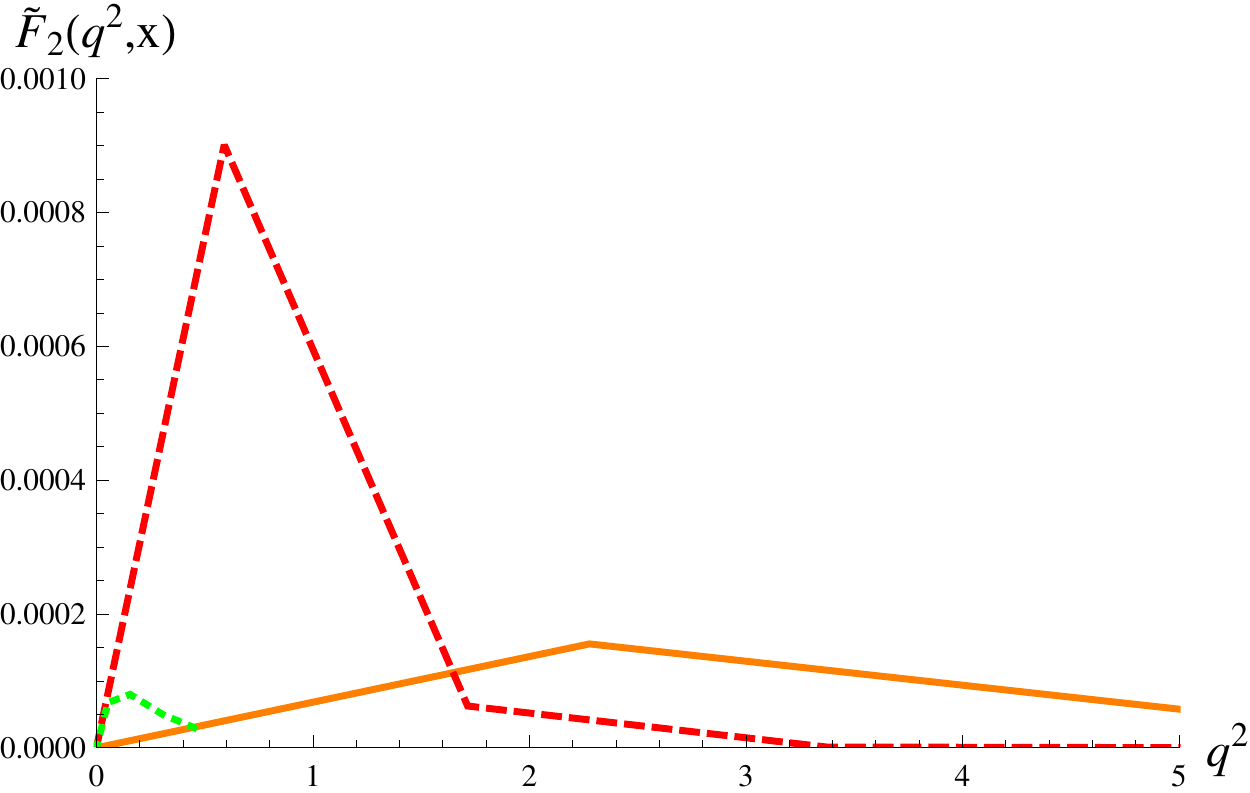}
\end{center}
\caption{Structure functions $F_{1,2}(q^2)$ for $x=0.3$ (orange, solid), $x=0.1$ (red, dashed) and $x=0.01$ (green, dotted).}
\label{fig:F12q2}
\end{figure}

\begin{figure}[h]
\begin{center}
\includegraphics[width=.38 \textwidth]{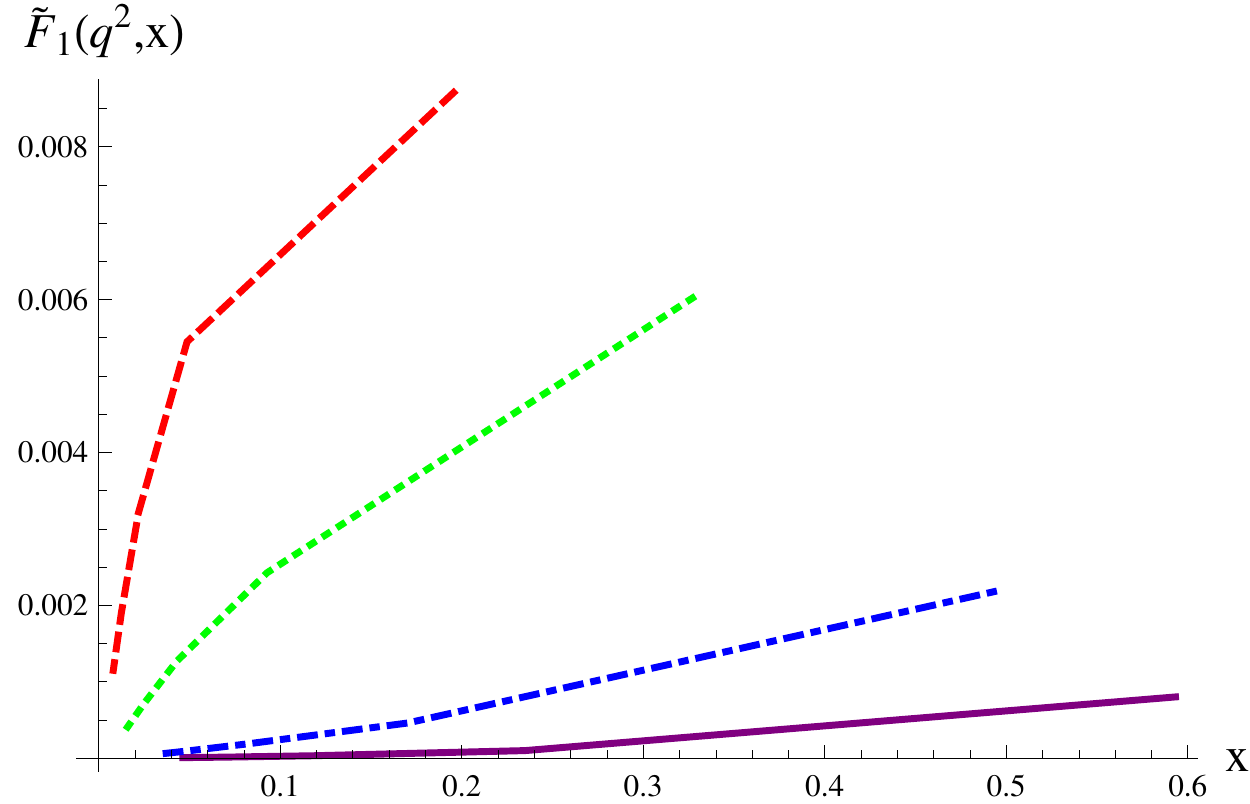}
\includegraphics[width=.38 \textwidth]{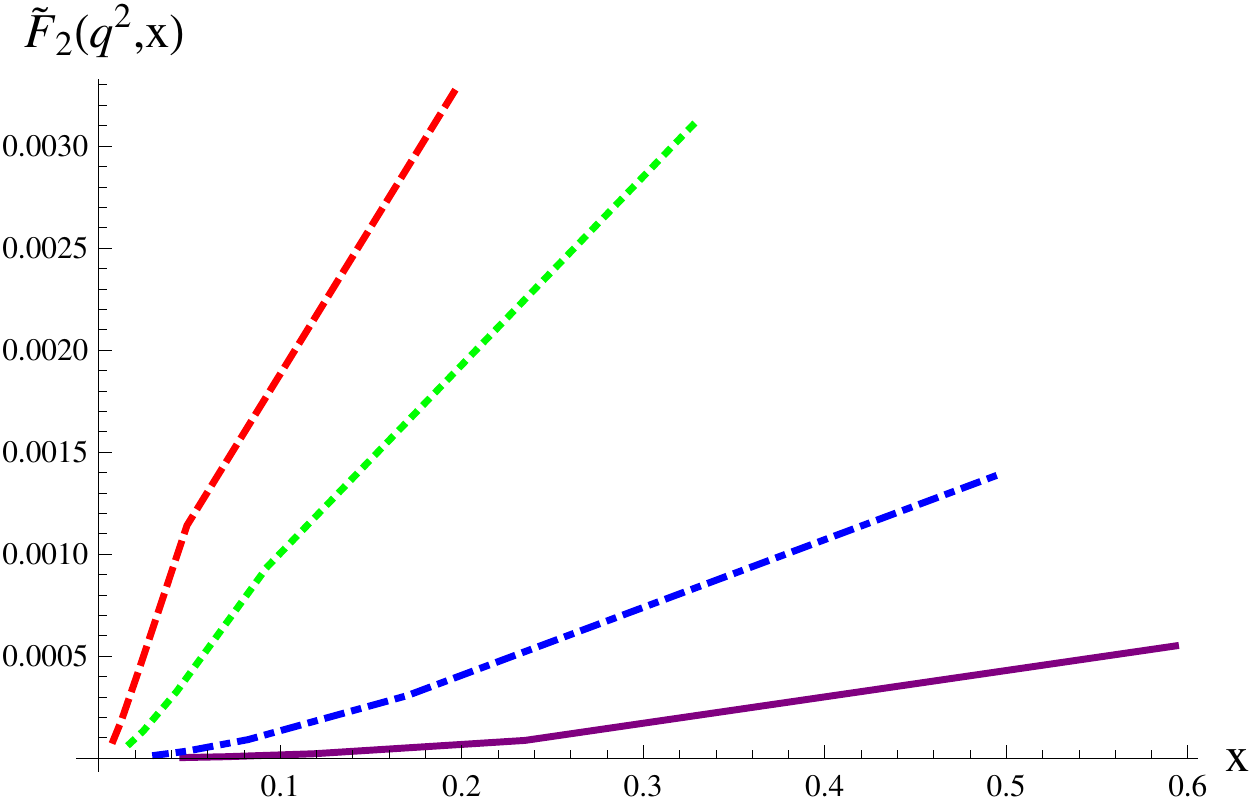}
\end{center}
\caption{Structure functions $\tilde F_{1,2}(x)$ for $q^2=3 (\text{GeV})^2$ (purple, solid), $q^2=2  (\text{GeV})^2$ (blue, dotdashed), $q^2=1  (\text{GeV})^2$ (green, dotted) and  $q^2=0.5 (\text{GeV})^2$ (red, dashed).}
\label{fig:F12x}
\end{figure}

\begin{figure}[h]
\begin{center}
\includegraphics[width=.38 \textwidth]{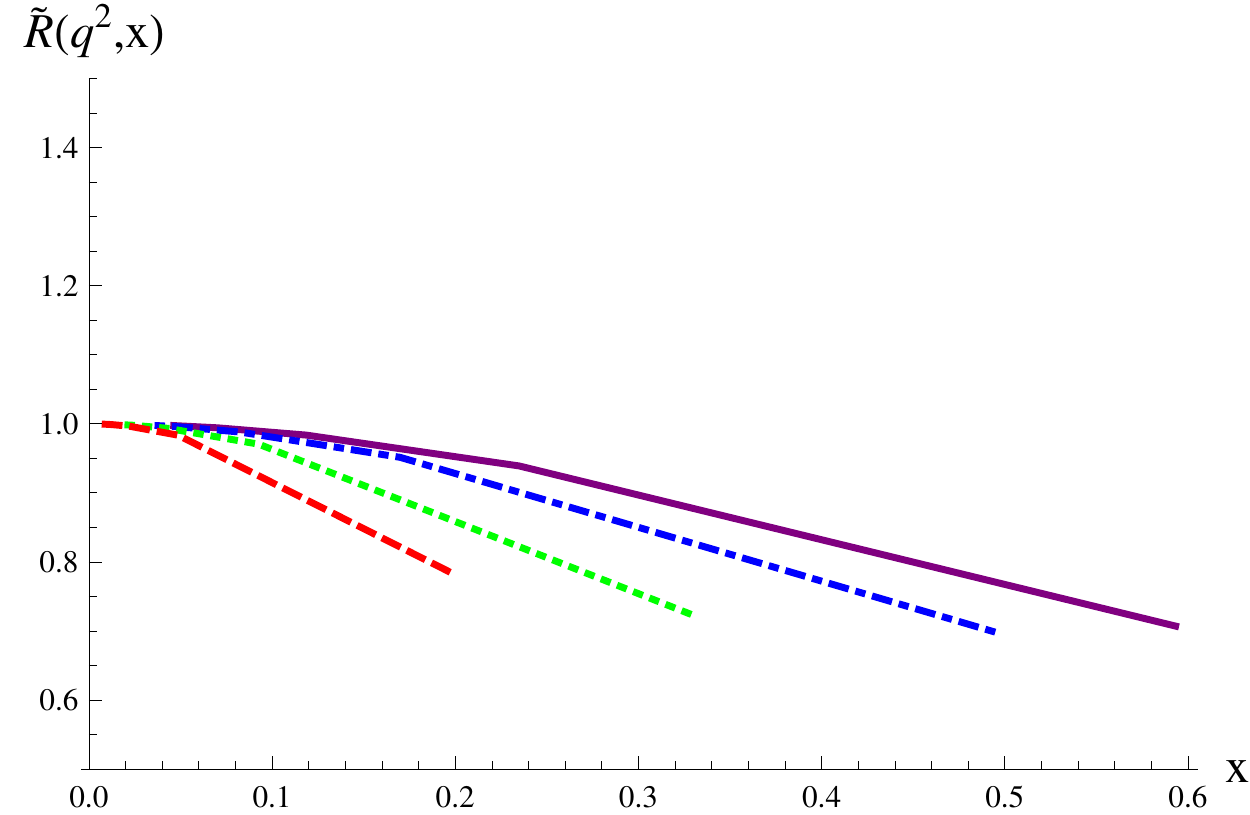}
\end{center}
\caption{Callan-Gross ratio $R_{\text{CG}}(x)$ for  $q^2=3 (\text{GeV})^2$ (purple, solid), $q^2=2  (\text{GeV})^2$ (blue, dotdashed), $q^2=1  (\text{GeV})^2$ (green, dotted) and  $q^2=0.5 (\text{GeV})^2$ (red, dashed).}
\label{fig:CGrel}
\end{figure}

We plot in figures \ref{fig:F12q2} and \ref{fig:F12x} the structure functions obtained from (\ref{naiveF12}) as
a function of $q^2$ and $x$. We also demonstrate the violation of the Callan-Gross relation at intermediate
values of $x$ in \ref{fig:CGrel}.

\begin{figure}[h]
\begin{center}
\includegraphics[width=.38 \textwidth]{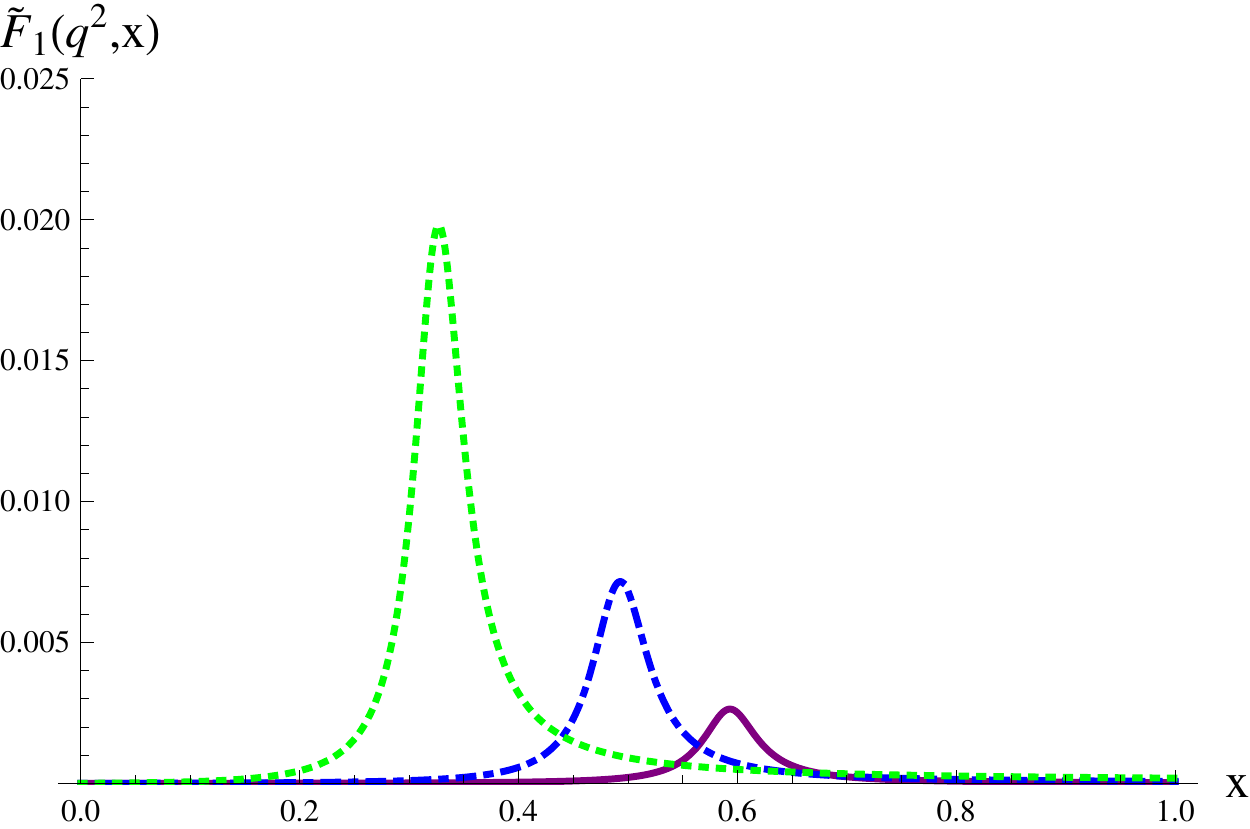}
\includegraphics[width=.38 \textwidth]{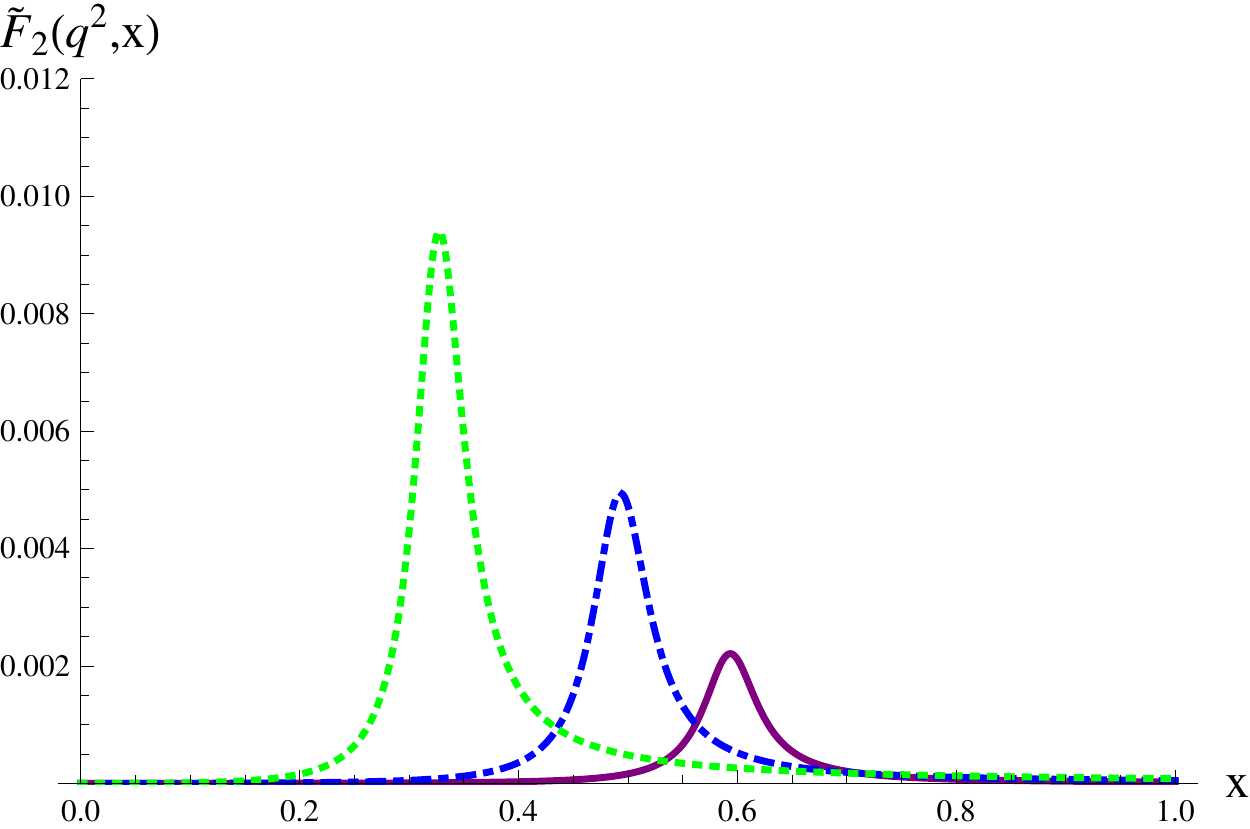}
\end{center}
\caption{Structure functions $F_{1,2}(x)$ for $q^2=3 (\text{GeV})^2$ (purple, solid),
$q^2=2  (\text{GeV})^2$ (blue, dotdashed), $q^2=1  (\text{GeV})^2$ (green, dotted) near the first
 negative parity resonance $B_1$}
\label{fig:ResF1F2vsx}
\end{figure}

\subsubsection{A realistic approach near a resonance peak}

The results  shown in figures \ref{fig:F12q2} and \ref{fig:F12x} were obtained
using a the naive approximation (\ref{approxDelta}). Alternatively, if we are only interested in the region
of $q^2$ where a resonance is produced we can approximate the Delta
distribution by a Lorentzian function \cite{Carlson:1998gf} :
\beqa
\delta [ m_{B_{X}}^2 -s ] \approx \frac{\Gamma_{B_{X} }}{ 4 \pi m_{B_{X}} }
\left [ (\sqrt{s} - m_{B_{X}})^2 + \frac{\Gamma_{B_{X}}^2}{4}  \right ]^{-1} \, ,
\eeqa
where $\Gamma_{B_{X}}$ is the decay width of the resonance $B_{X}$.
Identifying the first negative parity baryonic resonance $B_1$ with the experimentally observed $S_{11}(1535)$
and using the decay width $\Gamma_{B_1} = 150 \, {\rm Mev}$, estimated from experimental data
in  \cite{Beringer:1900zz}, we obtain the  results for the structure functions shown in Figure \ref{fig:ResF1F2vsx}.
Note that the structure functions have improved by an order of magnitude. Unfortunately, we cannot follow this procedure
for the higher resonances because there are no experimental results available for the decay widths.

The results for the proton structure functions obtained in this paper represent
only a small fraction of possible final states, namely single final states
with spin $1/2$ and negative parity.  If we include the contribution from final states with positive parity
\cite{Bayona:2011xj} as well as final states with higher spin\footnote{Usually one expects a high contribution
coming from the production of $\Delta$ resonances. See \cite{arXiv:0904.3710} for $\Delta$ resonances
 in the Sakai-Sugimoto model and \cite{Park:2008sp} for higher spin resonances.} and pion production relevant
in this kinematical regime, we should get a better/more complete picture of the proton structure functions and significantly improve the
comparison with experimental data.

\section{Conclusions and Outlook}
In this article we have presented a treatment of non-elastic proton electromagnetic scattering for the special case when baryonic resonances of negative parity
are produced as the single-particle final state of the scattering process. We have in turn applied the Sakai-Sugimoto model of holographic baryons in the large $\lambda$ limit
to compute the relevant form factors and proton structure functions. Our numerical results show good agreement with available experimental data. One should, however, keep in mind the limitations of the (holographic) description of baryons
in large-$N_c$ QCD \cite{Witten:1998xy}, which fully apply to the non-relativistic (large $\lambda$) model discussed herein as well.
It would be very interesting to calculate $1/\lambda$- and other corrections to the current model and to study other scattering processes within the Sakai-Sugimoto model. Finally,
it would be fruitful to investigate baryons and their resonance production in more recent holographic models, e.g., \cite{Kuperstein:2008cq, Dymarsky:2009cm, Ihl:2012at}.
We leave this for future work.

\section*{Acknowledgements}

The work of M.I. is supported by an IRCSET postdoctoral fellowship. A.B. is supported by the STFC Rolling Grant ST/G000433/1. The work of H.B., N.B. and M.T. are partially supported by CAPES and CNPq (Brazilian funding agencies). M.T. thanks Gilberto Ramalho for clarifications on EBAC results on bare helicity amplitudes. 

\appendix

\section{Some frames in inelastic scattering}

\subsection{The Breit frame}

Consider the scattering between a virtual photon and a hadron  in the hadron rest frame. After two rotations we can set the spatial momentum of the photon to the $x^3$ direction so
\beqa
p^\mu &=& (m_B , 0 , 0 ,0 ) \cr
q^\mu &=& (q_0, 0 , 0 , q_3 ) \, ,
\eeqa
and we choose $q_3 > 0$.
The virtuality and Bjorken variable are in this frame given by
\beqa
Q^2 = q_3^2 - q_0^2   \quad , \quad x = - \frac{Q^2}{2 m_B q_0} \, .
\eeqa

Now we perform a boost in the $x^3$ direction so that
\beqa
p'^\mu &=& (\gamma m_B , 0 , 0 , - \beta \gamma m_B ) \cr
q'^\mu &=& (\gamma q_0 - \beta \gamma q_3 , 0 , 0 , - \beta \gamma q_0 + \gamma q_3 ) \,.
\eeqa

The Breit Frame is defined by the condition $q'_0 =0$ so that
\beqa
\beta = \frac{q_0}{q_3}= \frac{q_0}{\sqrt{q_0^2 + Q^2}} \quad , \quad \gamma = \frac{\sqrt{q_0^2 + Q^2}}{Q} \quad , \quad q'_3 = Q  \,,
\eeqa
and we arrive to
\beqa
p'^\mu &=& (\sqrt{m_B^2 + p^2} , 0 , 0 , p ) \cr
q'^\mu &=& (0 , 0 , 0 , Q ) \,,
\eeqa
with
\beqa
p = - \frac{Q}{2x} \,.
\eeqa

\subsection{The resonant rest frame}

In the resonant frame we have
\beqa
p^\mu &=& (E_R \, , \,  - \vec{q}) \, , \cr
q^\mu &=& (m_{B_X} - E_R \, , \,  \vec{q}_R ) \,, \cr
(p + q)^\mu &=& ( m_{B_X} \, , \, 0 ) \,.
\eeqa
We can write the energy and the momentum squared in terms of the squared masses and virtuality
\beqa
E_R &=& \frac{1}{2 m_{B_X}} \left [ q^2 + m_{B_X}^2 + m_B^2 \right ] \cr
|\vec{q}_R|^2 &=& (E_R - m_B)(E_R + m_B) \, .
\eeqa

\section{Expansions at large $\lambda$}

The relevant large $\lambda$ expansions for the non-elastic case are given by
\beqa
q^2 &\sim& {\cal O}(1) \quad , \quad m_B  \sim {\cal O} (\lambda N_c) \quad \, ,\cr
m_{B_{\scst  X}} &=& m_B + \frac{2}{\sqrt{6}} n_{\scst  X} M_{\scst  KK} = m_B \left [ 1 +  {\cal O} \left ( \frac{1}{\lambda N_c } \right ) \right ] \, , \cr
x &=& \left (\frac{\sqrt{6}}{4} \right ) \frac{q^2}{m_B M_{\scst  KK}  n_{\scst  X}}  \left [ 1 +  {\cal O} \left ( \frac{1}{\lambda N_c } \right ) \right ] = {\cal O}  \left ( \frac{1}{\lambda N_c } \right ) \, , \cr
E &=& m_B \sqrt{ 1 + \frac23 \frac{ n_{\scst  X}^2 M_{\scst  KK}^2 }{q^2} }
 \left [ 1 + {\cal O} \left ( \frac{1}{\lambda N_c} \right ) \right ] \, , \cr
 \left ( \frac{M_0}{3} \right )  \langle \rho^2 \rangle &=&
\frac{g_{I=1}}{4 m_B} = {\cal O}(N_c) \quad , \quad
\alpha =  1 + {\cal O} \left ( \frac{1}{ \lambda N_c} \right ) \, , \cr
\beta  &=&  \frac{1}{2 m_B }  \left [ 1  +  {\cal O} \left ( \frac{1}{ \lambda N_c} \right )  \right ]
= {\cal O} \left ( \frac{1}{ \lambda N_c} \right )\, ,  \cr
\hat \alpha &=& \left ( \frac{f}{f_x} \right ) \left ( \frac{1}{2E} \right ) \left [ \frac{f_x^2}{f^2} + 1 - 2x \right ]
= \frac{1}{E} \left [ 1 + {\cal O} \left ( \frac{1}{ \lambda N_c} \right ) \right ] \, , \cr
\hat \beta &=&  \frac{2 x}{q^2} \xi = \left ( \frac{2 x}{q^2} \right ) \frac{m_B}{E} \left [ 1 + {\cal O} \left ( \frac{1}{ \lambda N_c} \right ) \right ]
= {\cal O} \left ( \frac{1}{ \lambda N_c} \right ) \, , \cr
\xi &=& \frac{m_B}{E}  +  {\cal O} \left ( \frac{1}{ \lambda N_c} \right ) \quad , \quad
\alpha^2 + \beta^2 q^2 = 1 + {\cal O} \left ( \frac{1}{ \lambda N_c} \right ) \, , \cr
\xi \alpha + \beta \alpha \frac{q^2}{4 M_0} &=& \frac{m_B}{E} + {\cal O} \left ( \frac{1}{ \lambda N_c} \right )
\, , \cr
- \frac{1}{\kappa_B} \left ( \beta \xi - \frac{\alpha^2}{4 M_0} \right ) &=&
\frac{g_{I=0}}{2} -\frac{m_B}{E}  + {\cal O} \left ( \frac{1}{ \lambda N_c} \right ) \cr
\xi \alpha + \beta \alpha q^2  \left ( \frac{M_0}{3} \right )  \langle \rho^2 \rangle  &=& \frac{m_B}{E} + {\cal O} \left ( \frac{1}{\lambda} \right ) \, , \cr
- \frac{1}{\kappa_B} \left [ \beta \xi - \alpha^2 \left ( \frac{M_0}{3} \right )  \langle \rho^2 \rangle \right ]
&=& \frac{g_{I=1}}{2} \left [ 1 + {\cal O} \left ( \frac{1}{ \lambda N_c} \right )\right] \,, \cr
\hat \alpha^2 + \hat \beta^2 q^2 &=&  \frac{4 x^2}{q^2} \left [ 1 + {\cal O} \left ( \frac{1}{ \lambda N_c} \right ) \right ]  \, , \cr
x \left ( \frac{q^2}{2M_0} \right ) \hat \beta \alpha &=& \frac{x^2}{E} g_{I=0} \left [ 1 + {\cal O} \left ( \frac{1}{ \lambda N_c} \right ) \right ]  \, , \cr
x \left ( \frac{1}{2 M_0 \kappa_B} \right ) \hat \alpha \alpha &=& \frac{x}{E} g_{I=0} \left [ 1 + {\cal O} \left ( \frac{1}{ \lambda N_c} \right ) \right ] \, , \cr
2 x \left [ \frac{M_0}{3} \langle \rho^2 \rangle  \hat \beta \alpha q^2  - \hat \alpha \xi  \right ] &=&  \frac{x^2}{E} g_{I=1} \left [ 1 + {\cal O} \left ( \frac{1}{N_c} \right ) \right ]  \, , \cr
2 x \left ( \frac{1}{\kappa_B} \right ) \left [ \frac{M_0}{3} \langle \rho^2 \rangle  \hat \alpha \alpha  + \hat \beta \xi  \right ] &=& \frac{x}{E} g_{I=1}
\left [ 1 + {\cal O} \left ( \frac{1}{N_c} \right ) \right ] \, .
\eeqa


\begin{thebibliography}{19}

\bibitem{Maldacena:1997re}
  J.~M.~Maldacena,
  ``The Large N limit of superconformal field theories and supergravity,''
  Adv.\ Theor.\ Math.\ Phys.\  {\bf 2}, 231 (1998)
  [hep-th/9711200].
\bibitem{hep-th/0610135}
  H.~Boschi-Filho and N.~R.~F.~Braga,
  ``AdS/CFT correspondence and strong interactions,''
  PoSIC\ {\bf 2006}, 035  (2006)
  [hep-th/0610135].
\bibitem{arXiv:0908.0312}
  J.~Erlich,
  ``How Well Does AdS/QCD Describe QCD?,''
  Int.\ J.\ Mod.\ Phys.\ A\ {\bf 25}, 411  (2010)
  [arXiv:0908.0312 [hep-ph]].
\bibitem{arXiv:1001.1978}
  S.~J.~Brodsky and G.~ de Teramond,
  ``AdS/QCD and Light Front Holography: A New Approximation to QCD,''
  Chin.\ Phys.\ C\ {\bf 34}, 1  (2010)
  [arXiv:1001.1978 [hep-ph]].
\bibitem{arXiv:1006.5461}
  U.~Gursoy, E.~Kiritsis, L.~Mazzanti, G.~Michalogiorgakis and F.~Nitti,
 ``Improved Holographic QCD,''
  Lect.\ Notes Phys.\ \ {\bf 828}, 79  (2011)
  [arXiv:1006.5461 [hep-th]].
\bibitem{Peeters:2007ab}
  K.~Peeters, M.~Zamaklar,
  ``The String/gauge theory correspondence in QCD,''
  Eur.\ Phys.\ J.\ ST {\bf 152}, 113-138 (2007).
  [arXiv:0708.1502 [hep-ph]].
\bibitem{Erdmenger:2007cm}
  J.~Erdmenger, N.~Evans, I.~Kirsch, E.~Threlfall,
  ``Mesons in Gauge/Gravity Duals - A Review,''
  Eur.\ Phys.\ J.\  {\bf A35}, 81-133 (2008).
  [arXiv:0711.4467 [hep-th]].
\bibitem{Gubser:2009md}
  S.~S.~Gubser, A.~Karch,
  ``From gauge-string duality to strong interactions: A Pedestrian's Guide,''
  Ann.\ Rev.\ Nucl.\ Part.\ Sci.\  {\bf 59}, 145-168 (2009).
 [arXiv:0901.0935 [hep-th]].
\bibitem{arXiv:1101.0618}
  J.~Casalderrey-Solana, H.~Liu, D.~Mateos, K.~Rajagopal and U.~A.~Wiedemann,
  ``Gauge/String Duality, Hot QCD and Heavy Ion Collisions,''
  arXiv:1101.0618 [hep-th].
\bibitem{Sakai:2004cn}
  T.~Sakai and S.~Sugimoto,
  ``Low energy hadron physics in holographic QCD,''
  Prog.\ Theor.\ Phys.\  {\bf 113}, 843 (2005)
  [arXiv:hep-th/0412141].
\bibitem{Sakai:2005yt}
  T.~Sakai and S.~Sugimoto,
  ``More on a holographic dual of QCD,''
  Prog.\ Theor.\ Phys.\  {\bf 114}, 1083 (2005)
  [arXiv:hep-th/0507073].
\bibitem{Bayona:2011xj}
  C.~A.~B.~Bayona, H.~Boschi-Filho, N.~R.~F.~Braga, M.~Ihl and M.~A.~C.~Torres,
  ``Generalized baryon form factors and proton structure functions in the Sakai-Sugimoto model,''
Nucl. Phys. B. {\bf 866}, 124 (2013)
arXiv:1112.1439 [hep-ph].
\bibitem{Aznauryan:2011ub}
  I.~Aznauryan, V.~D.~Burkert, T.~-S.~H.~Lee and V.~I.~Mokeev,
  ``Results from the N* program at JLab,''
  J.\ Phys.\ Conf.\ Ser.\  {\bf 299}, 012008 (2011)
  [arXiv:1102.0597 [nucl-ex]].
\bibitem{Hong:2007ay}
  D.~K.~Hong, M.~Rho, H.~U.~Yee and P.~Yi,
  ``Dynamics of baryons from string theory and vector dominance,''
  JHEP {\bf 0709}, 063 (2007)
  [arXiv:0705.2632 [hep-th]].
 \bibitem{Hashimoto:2008zw}
  K.~Hashimoto, T.~Sakai, S.~Sugimoto,
  ``Holographic Baryons: Static Properties and Form Factors from Gauge/String Duality,''
  Prog.\ Theor.\ Phys.\  {\bf 120}, 1093-1137 (2008).
  [arXiv:0806.3122 [hep-th]].
\bibitem{Kim:2008pw}
  K.~-Y.~Kim and I.~Zahed,
  ``Electromagnetic Baryon Form Factors from Holographic QCD,''
  JHEP {\bf 0809}, 007 (2008)
  [arXiv:0807.0033 [hep-th]].
\bibitem{arXiv:0903.4818}
  Z.~Abidin and C.~E.~Carlson,
  ``Nucleon electromagnetic and gravitational form factors from holography,''
  Phys.\ Rev.\ D\ {\bf 79}, 115003  (2009)
  [arXiv:0903.4818 [hep-ph]].
\bibitem{arXiv:0904.3710}
  H.~R.~Grigoryan, T.~-S.~H.~Lee and H.~-U.~Yee,
  ``Electromagnetic Nucleon-to-Delta Transition in Holographic QCD,''
  Phys.\ Rev.\ D\ {\bf 80}, 055006  (2009)
  [arXiv:0904.3710 [hep-ph]].
\bibitem{Park:2008sp}
  J.~Park and P.~Yi,
  ``A Holographic QCD and Excited Baryons from String Theory,''
  JHEP {\bf 0806}, 011 (2008)
  [arXiv:0804.2926 [hep-th]].

\bibitem{arXiv:0904.3731}
  H.~C.~Ahn, D.~K.~Hong, C.~Park and S.~Siwach,
  ``Spin 3/2 Baryons and Form Factors in AdS/QCD,''
  Phys.\ Rev.\ D\ {\bf 80}, 054001  (2009)
  [arXiv:0904.3731 [hep-ph]].
\bibitem{Polchinski:2002jw}
  J.~Polchinski, M.~J.~Strassler,
  ``Deep inelastic scattering and gauge / string duality,''
  JHEP {\bf 0305}, 012 (2003).
  [hep-th/0209211].
\bibitem{BallonBayona:2007qr}
  C.~A.~Ballon Bayona, H.~Boschi-Filho, N.~R.~F.~Braga,
  ``Deep inelastic scattering from gauge string duality in the soft wall model,''
  JHEP {\bf 0803}, 064 (2008).
  [arXiv:0711.0221 [hep-th]].
\bibitem{BallonBayona:2008zi}
  C.~A.~Ballon Bayona, H.~Boschi-Filho, N.~R.~F.~Braga,
  ``Deep inelastic scattering from gauge string duality in D3-D7 brane model,''
  JHEP {\bf 0809}, 114 (2008).
  [arXiv:0807.1917 [hep-th]].
\bibitem{Pire:2008zf}
  B.~Pire, C.~Roiesnel, L.~Szymanowski, S.~Wallon,
  ``On AdS/QCD correspondence and the partonic picture of deep inelastic scattering,''
  Phys.\ Lett.\  {\bf B670}, 84-90 (2008).
  [arXiv:0805.4346 [hep-ph]].
\bibitem{BallonBayona:2010ae}
  C.~A.~Ballon Bayona, H.~Boschi-Filho, N.~R.~F.~Braga, M.~A.~C.~Torres,
  ``Deep inelastic scattering for vector mesons in holographic D4-D8 model,''
  JHEP {\bf 1010}, 055 (2010).
  [arXiv:1007.2448 [hep-th]].
\bibitem{Brower:2006ea}
  R.~C.~Brower, J.~Polchinski, M.~J.~Strassler, C.~-ITan,
  ``The Pomeron and gauge/string duality,''
  JHEP {\bf 0712}, 005 (2007).
  [hep-th/0603115].
\bibitem{Hatta:2007he}
  Y.~Hatta, E.~Iancu, A.~H.~Mueller,
  ``Deep inelastic scattering at strong coupling from gauge/string duality: The Saturation line,''
  JHEP {\bf 0801}, 026 (2008).
  [arXiv:0710.2148 [hep-th]].
\bibitem{BallonBayona:2007rs}
  C.~A.~Ballon Bayona, H.~Boschi-Filho, N.~R.~F.~Braga,
  ``Deep inelastic structure functions from supergravity at small x,''
  JHEP {\bf 0810}, 088 (2008).
  [arXiv:0712.3530 [hep-th]].
\bibitem{arXiv:0804.1562}
  L.~Cornalba and M.~S.~Costa,
  ``Saturation in Deep Inelastic Scattering from AdS/CFT,''
  Phys.\ Rev.\ D\ {\bf 78}, 096010  (2008)
  [arXiv:0804.1562 [hep-ph]].
\bibitem{Cornalba:2009ax}
  L.~Cornalba, M.~S.~Costa, J.~Penedones,
  ``Deep Inelastic Scattering in Conformal QCD,''
  JHEP {\bf 1003}, 133 (2010).
  [arXiv:0911.0043 [hep-th]].
\bibitem{Brower:2010wf}
  R.~C.~Brower, M.~Djuric, I.~Sarcevic, C.~-ITan,
  ``String-Gauge Dual Description of Deep Inelastic Scattering at Small-$x$,''
  JHEP {\bf 1011}, 051 (2010).
  [arXiv:1007.2259 [hep-ph]].
\bibitem{Hatta:2007cs}
  Y.~Hatta, E.~Iancu and A.~H.~Mueller,
  ``Deep inelastic scattering off a N=4 SYM plasma at strong coupling,''
  JHEP {\bf 0801}, 063 (2008)
  [arXiv:0710.5297 [hep-th]].
\bibitem{Bayona:2009qe}
  C.~A.~B.~Bayona, H.~Boschi-Filho and N.~R.~F.~Braga,
  ``Deep inelastic scattering off a plasma with flavour from D3-D7 brane model,''
  Phys.\ Rev.\ D {\bf 81}, 086003 (2010)
  [arXiv:0912.0231 [hep-th]].
\bibitem{Iancu:2009py}
  E.~Iancu and A.~H.~Mueller,
  ``Light-like mesons and deep inelastic scattering in finite-temperature AdS/CFT with flavor,''
  JHEP {\bf 1002}, 023 (2010)
  [arXiv:0912.2238 [hep-th]].
\bibitem{Bu:2011my}
  Y.~Y.~Bu and J.~M.~Yang,
  ``Structure function of holographic quark-gluon plasma: Sakai-Sugimoto model versus its non-critical version,''
  Phys.\ Rev.\ D {\bf 84}, 106004 (2011)
  [arXiv:1109.4283 [hep-th]].
\bibitem{Aznauryan:2008us}
  I.~G.~Aznauryan, V.~D.~Burkert and T.~-S.~H.~Lee,
  ``On the definitions of the gamma* N $\to$ N* helicity amplitudes,''
  arXiv:0810.0997 [nucl-th].
\bibitem{Carlson:1998gf}
  C.~E.~Carlson and N.~C.~Mukhopadhyay,
  ``Bloom-Gilman duality in the resonance spin structure functions,''
  Phys.\ Rev.\ D {\bf 58}, 094029 (1998)
  [hep-ph/9801205].
\bibitem{Stoler:1993yk}
  P.~Stoler,
  ``Baryon form-factors at high Q**2 and the transition to perturbative QCD,''
  Phys.\ Rept.\  {\bf 226}, 103 (1993).
\bibitem{Manohar:1992tz}
  A.~V.~Manohar,
  ``An introduction to spin dependent deep inelastic scattering,''
  arXiv:hep-ph/9204208.
\bibitem{Witten:1998qj}
  E.~Witten,
  ``Anti-de Sitter space and holography,''
  Adv.\ Theor.\ Math.\ Phys.\  {\bf 2}, 253 (1998)
  [arXiv:hep-th/9802150].
 \bibitem{Hata:2007mb}
  H.~Hata, T.~Sakai, S.~Sugimoto, S.~Yamato,
  ``Baryons from instantons in holographic QCD,''
  Prog.\ Theor.\ Phys.\  {\bf 117}, 1157 (2007).
  [hep-th/0701280 [hep-th]].
\bibitem{BoschiFilho:2011hn}
  H.~Boschi-Filho, N.~R.~F.~Braga, M.~Ihl and M.~A.~C.~Torres,
  ``Relativistic baryons in the Skyrme model revisited,''
  Phys.\ Rev.\ D {\bf 85}, 085013 (2012)
  [arXiv:1111.2287 [hep-th]].
\bibitem{BallonBayona:2009ar}
  C.~A.~Ballon Bayona, H.~Boschi-Filho, N.~R.~F.~Braga and M.~A.~C.~Torres,
  ``Form factors of vector and axial-vector mesons in holographic D4-D8 model,''
  JHEP {\bf 1001}, 052 (2010)
  [arXiv:0911.0023 [hep-th]].
\bibitem{Bayona:2010bg}
  C.~A.~B.~Bayona, H.~Boschi-Filho, M.~Ihl and M.~A.~C.~Torres,
  ``Pion and Vector Meson Form Factors in the Kuperstein-Sonnenschein holographic model,''
  JHEP {\bf 1008}, 122 (2010)
  [arXiv:1006.2363 [hep-th]].
\bibitem{Xie:2008ts}
  J.~-j.~Xie, C.~Wilkin and B.~-s.~Zou,
  ``On the Coupling Constant for N*(1535)N(rho),''
  Phys.\ Rev.\ C {\bf 77}, 058202 (2008)
  [arXiv:0802.2802 [nucl-th]].
\bibitem{Aznauryan:2009mx}
  I.~G.~Aznauryan {\it et al.}  [CLAS Collaboration],
  ``Electroexcitation of nucleon resonances from CLAS data on single pion electroproduction,''
  Phys.\ Rev.\ C {\bf 80}, 055203 (2009)
  [arXiv:0909.2349 [nucl-ex]].
\bibitem{Cherman:2009gb}
  A.~Cherman, T.~D.~Cohen and M.~Nielsen,
  ``Model Independent Tests of Skyrmions and Their Holographic Cousins,''
  Phys.\ Rev.\ Lett.\  {\bf 103}, 022001 (2009)
  [arXiv:0903.2662 [hep-ph]].
  \bibitem{Matsuyama:2006rp} 
  A.~Matsuyama, T.~Sato and T.~-S.~H.~Lee,
  Phys.\ Rept.\  {\bf 439}, 193 (2007)
  [nucl-th/0608051].
\bibitem{JuliaDiaz:2009ww} 
  B.~Julia-Diaz, H.~Kamano, T.~-S.~H.~Lee, A.~Matsuyama, T.~Sato and N.~Suzuki,
  Phys.\ Rev.\ C {\bf 80}, 025207 (2009)
  [arXiv:0904.1918 [nucl-th]].

\bibitem{Ramalho:2011ae} 
  G.~Ramalho and M.~T.~Pena,
  Phys.\ Rev.\ D {\bf 84}, 033007 (2011)
  [arXiv:1105.2223 [hep-ph]].
\bibitem{Beringer:1900zz} 
  J.~Beringer {\it et al.}  [Particle Data Group Collaboration],
  Phys.\ Rev.\ D {\bf 86}, 010001 (2012).

 
\bibitem{Witten:1998xy}
  E.~Witten,
  ``Baryons and branes in anti-de Sitter space,''
  JHEP {\bf 9807}, 006 (1998)
  [arXiv:hep-th/9805112].
 \bibitem{Kuperstein:2008cq}
  S.~Kuperstein and J.~Sonnenschein,
  ``A New Holographic Model of Chiral Symmetry Breaking,''
  JHEP {\bf 0809}, 012 (2008)
  [arXiv:0807.2897 [hep-th]].
\bibitem{Dymarsky:2009cm}
  A.~Dymarsky, S.~Kuperstein and J.~Sonnenschein,
  ``Chiral Symmetry Breaking with non-SUSY D7-branes in ISD backgrounds,''
  JHEP {\bf 0908}, 005 (2009)
  [arXiv:0904.0988 [hep-th]].
\bibitem{Ihl:2012at}
  M.~Ihl, A.~Kundu and S.~Kundu,
  ``Back-reaction of Non-supersymmetric Probes: Phase Transition and Stability,''
  arXiv:1208.2663 [hep-th].


  
\end{thebibliography}
\end{document}